\documentclass[10pt,preprint]{aastex}
\newcommand{\nh}{${\rm N_H}$}
\newcommand{\msun}{${\rm M_{\bigodot}}$}
\newcommand{\lsun}{${\rm L_{\bigodot}}$}
\newcommand{\lax}{${_<\atop^{\sim}}$}

\accepted{ApJ, 590, June 10}

\begin{document}

\title{The Far-Infrared Spectral Energy Distributions of X-ray-selected
Active Galaxies.\footnote{Based on observations with the Infrared Space
Observatory, which is an ESA project with instruments funded by ESA
Member States (especially the PI countries: France, Germany, the
Netherlands and the United Kingdom) and with the participation of ISAS
and NASA.}}

\author{Joanna K. Kuraszkiewicz, Belinda J. Wilkes}
\affil{Harvard-Smithsonian Center for Astrophysics,
60 Garden St, Cambridge, MA 02138 \\
{\it jkuraszkiewicz@cfa.harvard.edu, bwilkes@cfa.harvard.edu}}

\author{Eric, J. Hooper}
\affil{Department of Astronomy, University of Texas, Austin, TX 78712\\
{\it ehooper@astro.as.utexas.edu}}

\author{Kenneth Wood}
\affil{Department of Physics and Astronomy, University of St.
Andrews, St. Andrews, Fife, Scotland\\ 
{\it kw25@st-andrews.ac.uk}}

\author{Jon Bjorkman}
\affil{Ritter Observatory, Department of Physics and Astronomy, 
University of Toledo, Toledo, OH 43606\\
{\it jon@astro.utoledo.edu}}

\author{Kisha M. Delain}  
\affil{University of Minnesota, Department of Astronomy, Minneapolis, MN 55455\\
{\it kdelain@astro.umn.edu}}

\author{Barbara Whitney}
\affil{Space Science Institute, Boulder, CO 80303\\
{\it bwhitney@colorado.edu}}

\author{David H. Hughes}
\affil{Instituto National de Astrofisica, Optica y Electronica, Mexico\\ 
{\it dhughes@inaoep.mx}}

\author{Kim K. McLeod}
\affil{Wellesley College, Astronomy Department, Wellesley, MA 02481\\
{\it kmcleod@wellesley.edu}} 

\author{Martin S. Elvis}
\affil{Harvard-Smithsonian Center for Astrophysics,
60 Garden St, Cambridge, MA 02138\\
{\it melvis@cfa.harvard.edu}} 

\author{Chris D. Impey}
\affil{Steward Observatory, University of Arizona, Tucson, AZ 85721\\
{\it impey@as.arizona.edu}}

\author{Carol J. Lonsdale}
\affil{IPAC, Caltech 100-22, Pasadena, CA 91125\\
{\it cjl@ipac.caltech.edu}} 

\author{Matt A. Malkan}
\affil{UCLA, Astronomy Department, Los Angeles, CA 90095\\
{\it malkan@bonnie.astro.ucla.edu}}

\author{Jonathan C. McDowell}
\affil{Harvard-Smithsonian Center for Astrophysics, Cambridge, MA 02138\\
{\it jmcdowell@cfa.harvard.edu}}

\begin{abstract}

Hard X-ray selection is, arguably, the optimal method for defining a
representative sample of active galactic nuclei (AGN). Hard X-rays are
unbiased by the affects of obscuration and re-processing along the
line-of-sight intrinsic/external to the AGN which result in unknown
fractions of the population being missed from traditional
optical/soft-X-ray samples. We present the far-infrared (IR)
observations of 21 hard X-ray selected AGN from the HEAO-1 A2 sample
observed with ISO.  We characterize the far-infrared (IR) continua of
these X-ray selected AGN, compare them with those of various radio and
optically selected AGN samples and with models for an AGN-heated,
dusty disk.  The X-ray selected AGN show broad, warm IR continua
covering a wide temperature range ($\sim 20-1000$\,K in a thermal
emission scenario). Where a far-IR turnover is clearly observed, the
slopes are $< 2.5$ in all but three cases so that non-thermal emission
remains a possibility, although the presence of cooler dust resulting
in a turn-over at wavelengths longwards of the ISO range is considered
more likely. The sample also shows a wider range of optical/UV shapes
than the optically/radio-selected samples, extending to redder near-IR
colors.  The bluer objects are type 1 Seyferts, while the redder AGN
are mostly intermediate or type 2 Seyferts. This is consistent with a
modified unification model in which obscuration increases as we move
from a face-on towards more edge-on line-of-sight (l.o.s.) However, this
relation does not extend to the mid-infrared as the 25$\mu$m/60$\mu$m
ratios are similar in Seyferts with differing type and optical/UV
reddening.  The resulting limits on the column density of obscuring
material through which we are viewing the redder AGN (\nh~$\sim
10^{22}$ cm$^{-2}$) are inconsistent with standard optically thick
torus models (\nh~$\sim 10^{24}$ cm$^{-2}$) and simple unification
models.  Instead our results support more complex models in which the
amount of obscuring material increases with viewing angle and may be
clumpy.  Such a scenario, already suggested by differing
optical/near-IR spectroscopic and X-ray AGN classifications, allows
for different amounts of obscuration of the continuum emission in
different wavebands and of the broad emission line region which, in
turn, results in a mixture of behaviors for AGN with similar optical
emission line classifications. The resulting decrease in the optical
depth of the obscuring material also allows the AGN to heat more dust
at larger radial distances.  We show that an AGN-heated, flared, dusty
disk with mass $\sim 10^9$ \msun~and size$\sim$few hundred pc is able to
generate optical$-$far-IR spectral energy distributions (SEDs) which
reproduce the wide range of SEDs present in our sample with no need
for an additional starburst component to generate the long-wavelength,
cooler part of the IR continuum.

\end{abstract}

\section{Introduction}

Active galactic nuclei\index{active galactic nuclei} are among the
broadest waveband emission sources in nature, producing significant
flux over a span $> 9$ decades in frequency, from radio to X-rays and
beyond (Elvis et al. 1994).  The various emission mechanisms involved
are presumably ultimately powered by a central supermassive black hole
(Rees 1984).

A substantial fraction of the bolometric luminosity of many AGN
emerges in the infrared, from synchrotron radiation\index{synchrotron
radiation} and dust\index{dust emission}.  Which of these is the
principal emission mechanism is related to quasar type and is an open
question in many cases (Wilkes 1999a).  Non-thermal emission is
paramount in core dominated radio-loud quasars and blazars (Impey \&
Neugebauer 1988), although hot dust contributes in some cases
(Courvoisier 1998).  The non-thermal component is likely related to
radio and higher frequency synchrotron radiation (Chini et al. 1987),
providing information about the relativistic plasma and magnetic
fields associated with quasars.  Other AGN classes show evidence for a
predominant dust contribution (Edelson \& Malkan 1987), particularly
infrared-luminous radio-quiet quasars (Hughes et al. 1993), or a mix
of emission components (Haas et al. 1998).  Much of the dust emission
is due to heating by higher energy photons from the active nucleus,
and is therefore important for understanding the overall energy
balance. The AGN thermal component may be an orientation-independent
parameter, useful for examining unification hypotheses.

The nature of the foremost infrared emission source is ambiguous in
many AGN with sparsely sampled spectral energy
distributions\index{spectral energy distribution} (SEDs).  Dust with
smooth spatial and temperature distributions can mimic a power law
spectrum (Rees et al. 1969, Bollea \& Cavaliere 1976), particularly in
the absence of detailed measurements to reveal bumps from temperature
and density inhomogeneities.  Grain emissivity is characterized by a
Planck function multiplied by a power law factor $\propto \nu^{1 - 2}$
(Hildebrand 1983), so spectral slopes in the Rayleigh-Jeans region of
the coldest potential thermal component lie between $\alpha = 2$--4,
depending on the grain properties and optical depth.  Known
synchrotron emitters have relatively flat sub-mm power-law spectra
($\alpha \leq 1.1; f_{\nu} \propto \nu^{\alpha}$, Gear et al. 1994),
and radio sources generally have spectra flatter than the canonical
$\alpha = 2.5$ for a self-absorbed homogeneous synchrotron source
(O'Dea 1998).  Therefore, $\alpha = 2.5$ is a convenient partition to
distinguish thermal emission from standard non-thermal models, and
simple two-point spectral slopes and even lower limits to spectral
indices may reveal the dominant mechanism (Chini et al. 1987).
However, synchrotron models with a concave electron energy
distribution (de Kool \& Begelman 1989, Schlickeiser, Biermann \&
Crusius-W\"{a}tzel 1991), free-free absorption, or plasma suppression
(Schlickeiser \& Crusius 1989) can produce slopes steeper than $\alpha
= 2.5$.  While $\alpha = 4$ is observed in some milliarcsecond radio
knots (Matveyenko \& Witzel 1999), thermal models offer the most
consistent explanation for steep far-infrared (FIR) to mm slopes
(Hughes et al. 1993, Andreani, Franceschini \& Granato 1999).  A
thermal origin is considered to be the most likely explanation for
sub-mm/FIR slopes $\alpha > 2.5$ in the present work.

The unified models of AGN (e.g. Antonucci 1993) claim that the
difference between the broad-line AGN, such as Seyfert~1, and
narrow-line AGN, such as Seyfert~2, is due to orientation of a
circum-nuclear, dusty, torus-like structure. The dust absorbs the
emission from the central nucleus and reradiates it in the IR. The
first models of thermal IR emission in AGN included simple models as
in Barvainis (1987), where the optically thin (in IR) dust is
distributed smoothly in a disk-like configuration or as in Laor \&
Drain (1993) who considered a simple slab geometry. Sanders et
al. (1989) proposed a warped, dusty disk, extending into the host
galaxy. Recent models of the absorbing structure around the active
nucleus have invoked an axially-symmetric, torus-like geometry, which
is either compact (r$\le$ few pc) with large optical thickness
(\nh$\sim 10^{24}$ cm$^{-2}$ and $\tau \sim 1000$ in the UV) as in
Pier \& Krolik (1992) or extended (tens to hundred pc in diameter)
with moderate ($\tau =10-300$ in UV) optical depth as in Granato,
Danese, \& Francheschini (1997, see also: Granato \& Danese 1994,
Fadda et al. 1998 and Efstathiou \& Rowan-Robinson 1995). The extended
models have been confirmed by the observation of large ($\sim$ 100~pc)
extended dusty disk-like structures around NGC~4261 (Jaffe et
al. 1993), NGC~1068 (e.g. Young et al. 1996), while the moderate
optical depth was confirmed by the detection of broad near-IR
emission lines in optically narrow-lined AGN (Heisler \& De Robertis
1999), and by the weakness of the 10$\mu$m silicate absorption feature
in Seyfert 1 galaxies (Roche et al. 1991).
 
Two large, complementary ISO AGN observing programs have opened up new
wavelength windows in the FIR as well as improving the spatial
resolution and sampling at shorter wavelengths: the ISO European
Central Quasar Program (Haas et al. 1998, Haas et al. 2000); and the
NASA/ISO Key Project on AGN, discussed herein.  These observations
directly measure the FIR spectral slopes in low and moderate redshift
AGN and provide better constraints on the emission mechanisms
throughout the infrared region.  Some of the fundamental questions
being addressed with the new data include: the range of SEDs within
each AGN type, differences between one type and another, and
correlations with fluxes at other wavebands, and orientation
indicators.  This paper focuses on a random subset of the hard X-ray
selected AGN (Grossan 1992) in the HEAO catalog (Wood et
al. 1984). The advantage of hard X-ray selection over selection in
other wavebands is the lack of a strong bias against sources with
significant l.o.s. absorption.  Optical and soft X-ray samples, in
particular, are strongly affected and may miss a large fraction of the
AGN population as a whole (Masci et al. 1999, Webster et
al. 1995). The current sample thus provides an improved estimate of
the range of SEDs present in the AGN population as a whole as well as
an estimate of the fraction of the population missing from other
surveys.

\section{Observations and Data Analysis.}

\subsection{The Sample - Hard X-ray Selected AGN}

Infrared and X-ray data complement each other well and are important
for understanding the overall AGN energy balance.  Non-thermal
infrared emission is possibly connected to the X-ray emission, either
directly as part of a broad synchrotron component or as part of a
radio-infrared seed spectrum which Compton scatters to produce the
X-rays (see Wilkes 1999a for a review).  Infrared data from
dust-dominated sources reveal the level of ultra-violet (UV) and soft
X-ray radiation which has been reprocessed, and provide crude
estimates of the temperature and mass of the dust.

Most X-ray selected AGN to date have been found in soft X-ray bands
$<3.5$ keV (e.g. Stocke et al. 1991, Thomas et al. 1998, Beuermann et
al. 1999, Schwope et al. 2000).  Similar to optically selected
samples, soft X-ray samples are biased against obscured sources, in
this case due to absorption by the gas associated with the dust
responsible for the optical obscuration. The absorption cross-section
drops steeply with increasing energy (Zombeck 1990), so that hard
X-ray selection is much less affected by intervening material. Surveys
in hard X-rays arguably are the most efficient way to distinguish
between accreting and stellar sources (Fiore et al. 1999) as well as
the optimal method for defining a representative sample of AGN (Wilkes
1999b; as is the 2~micron near-IR selection - see Wilkes 2002).  A
comparison between the absorbed UV/soft X-ray flux and the far-IR
emission provides an estimate of the relative importance of accretion
and stellar power in AGN.

Hard X-ray samples (meaning heareafter selected in the 2-10~keV
band\footnote{Due to this definition Mkn~478, which has a very steep
soft X-ray spectrum, is included in our sample as being sufficiently
bright in the 2-10~keV band.}) have been difficult to obtain and are
limited in number of sources, depth, and redshift (although this is
changing rapidly with the advent of Chandra and XMM-Newton:
Hornschemeier et al. 2001, Brandt et al. 2001, Barger et al. 2001,
Tozzi et al. 2001, Wilkes et al. 2001, Hasinger et al. 2001).  One of
the best studied samples is that of Piccinotti et al. (1982), derived
from a large-area 2--10 keV survey using the A2 experiment on board
the HEAO 1 satellite (Giacconi et al. 1979, Marshall et al.1979, Boldt
1987). This sample contains 35 $z \leq 0.17$ AGN, mostly Seyfert 1s
(Kotilainen et al. 1992, Malizia et al. 1999).  Improvements in
analysis techniques pushed the flux limits deeper (Jahoda et
al. 1989), with a corresponding increase in the number of cataloged
sources.  We randomly selected 21 targets from an expanded compilation
of HEAO1-A2 AGN (Grossan 1992), 12 of which are also in the Piccinotti
sample.  We confirmed that our subset is representative of the whole
sample of HEAO AGN by comparing their optical and near-IR colors,
which have a $>$80\% chance of coming from the same population as the
full sample (Kolmogorov-Smirnov (K-S) test).  In this paper we analyze
these HEAO targets. Their redshift, coordinates and Seyfert type were
taken from NASA Extragalactic Database (NED)\footnote{The NASA/IPAC
Extragalactic Database (NED) is operated by the Jet Propulsion
Laboratory, California Institute of Technology, under contract with
the National Aeronautics and Space Administration.}  and are presented
in Table~\ref{tab:names}.

\subsection{ISO Observations \& Data Reduction}

The Key Project sample consists of 5--200~$\mu \rm{m}$ chopped and
rastered ISOPHOT (Lemke et al. 1996) observations of 73 AGN selected
to incorporate a wide range of AGN types and redshifts.  Eight
broadband ISOPHOT filters were employed: 4.85; 7.3; 11.5; 25; 60; 100;
135; and $200~\mu \rm{m}$. Most of the sources (53 of 72) were
observed in a rectangular chop mode, the point source detection
technique preferred at the beginning of the ISO mission.  Concerns
about calibrating and interpreting chopped measurements, particularly
at long wavelengths, led to a switch to small raster scans.  We
reobserved 18 of the chopped fields and switched 19 additional targets
from our original list to raster mode. The change in observing
strategy, combined with lower than expected instrumental sensitivity,
resulted in a halving of the originally planned sample.  However, data
from both observing modes are now available for a subset of the
targets, providing a cross-check and better information about
background variations from the raster maps.

The data are reduced using a combination of the (ISO-) PHOT
Interactive Analysis (PIA; Gabriel, Acosta-Pulido, \& Heinrichsen
1998) software plus custom scripts (Hooper et al. 1999a). Raster maps
have proven to be relatively reliable and are now considered
scientifically validated. Short wavelength, $\lambda \leq 25 \mu
\rm{m}$, chopped data obtained with the P1 and P2 detectors are also
fairly robust, but chopped measurements using the long-wavelength C1
and C2 arrays are still somewhat problematic (see the comparison with
IRAS fluxes in (Hooper et al. 1999a).  Some of the difficulty lies
with vignetting corrections, particularly for the C2 detector, in which
the source is centered at the convergence of 4 pixels (Hooper et
al. 1999a).  Detector drift may impart to the vignetting correction an
apparent dependence on chopper dwell time and brightness (Haas private
communication), an effect which is currently being investigated.
The ISO fluxes for subset of the HEAO1-A2 sample are presented in
Table~\ref{tab:ISO}.

\subsection{Spectral Energy Distributions}

The ISO points described above were combined with literature data
spanning from radio to hard-X-ray (see Table~\ref{tab:ebv_star} for
details) and complemented with our own near-IR photometry to generate
the most comprehensive SEDs possible.  We derived near-infrared
(JH$K_s$) photometry from images obtained on the Steward Observatory
61" telescope on Mt. Bigelow during the period June 1995 $-$ March
1996.  We used a 256$\times$256 NICMOS array camera with the pixel
scale set to $0\farcs9$.  To maximize on-source integration time and
to ensure proper monitoring of the sky, we took the images in a
$4\times4$ raster pattern for each filter.  Typical times at each
raster position varied from 1 $-$ 60 seconds, chosen to avoid
saturating the detector.  The 16 frames in each raster were used along
with a dark frame of the same exposure time to construct median sky
frames and normalized flats.  A clipping algorithm used during median
filtering ensured that the wings of the point sources were eliminated.
The reduced images were aligned and averaged to produce a final image
for each raster.  Elias photometric standard stars (Elias et al. 1982)
observed in the same manner provided aperture fluxes accurate to
5$-$10\% on nights with usable data.  The JHK photometry is presented
in Table~\ref{tab:photom_jhk}. In Mkn~1152 we include in the SED the
optical spectrophotometry carried out on October 07 1996, with the
FAST spectrograph on the 1.5~m Tillinghast telescope on Mt. Hopkins in
Arizona, using an $\sim$~3" aperture. To provide flux calibration, a
standard star was observed through the same aperture, at similar air
mass, immediately after the AGN observation. We reduced the data in a
standard manner using IRAF\footnote{IRAF (Image Reduction and Analysis
Facility) is distributed by the National Optical Astronomy
Observatories, which are operated by AURA, Inc., under cooperative
agreement with the National Science Foundation.}  (see Tokarz \& Roll
1997 for details) and then fit a continuum to the spectrum (using IRAF
task ``continuum''), which was next binned into broader wavelength
bands and included in the SED.

For comparison with the ISO data we also included observations from
the InfraRed Astronomical Satellite (IRAS). In one object
(PG~0804+761), however, the ISO chopped 
flux at 100~$\mu$m differed from the flux obtained with IRAS. We
decided to discard the ISO 100~$\mu$m flux in this object, as the C2
detector measurements in chopped mode may not be reliable (see
previous section), and included only the IRAS flux.  In a number of
objects (Ton~1542, MCG$-$6$-$30$-$15, IC~4329A, H~1419+480, E~1821+643,
Mkn~509, MR~2251$-$178) we also decided to omit the chopped 12~$\mu$m ISO
flux, as it differed from the IRAS flux considerably and did not
follow the overall SED shape.

A correction for galactic extinction was applied to each SED, which
was determined using the reddening values obtained from the literature
(see Table~\ref{tab:ebv_star} for details).  After correcting for
galactic extinction the data were blueshifted to the rest frame using
a cosmological model with $q_{o} = 0$ and $H_{o}$ = 50
kms$^{-1}$Mpc$^{-1}$.  No K-correction was applied since we were
working with complete SEDs. The contribution from the host galaxy was
subtracted using the method and template of Elvis et al. (1994) (based
on the Sbc galaxy model of Coleman, Wu and Weedman 1980) and
normalized by the host galaxy monochromatic luminosity in the H band
(or V band when H was not available) obtained from the literature (see
Table~\ref{tab:ebv_star} for details). If an AGN had more than one
data point in a single frequency bin, we calculated an average (in
$\log \nu f(\nu)$) of those observation points and included one data
point in the SED.  The final SEDs are displayed in
Figure{~\ref{fig:SEDs} in order of right-ascension. 

In order to better characterize the SEDs, we directly measured various
IR luminosities and slopes (presented in Table~\ref{tab:integrIR}) as
well as the optical, UV, X-ray and bolometric luminosities (presented
in Table~6).  The SEDs were linearly interpolated between the
observational points, while the EUV continuum was determined by
interpolating between the highest energy UV data point and the lowest
energy soft-X-ray point estimated from the observed flux and spectral
slope $\alpha_{x}$ (f$_{\nu} \propto \nu^{\alpha_x}$).  If the
soft-X-ray spectral slope was unknown the slope $\alpha_{x}=-1$ was
assumed.

We also plot a median AGN energy distribution in the UVOIR range
normalized to 1.5$\mu$m in Fig.~\ref{fig:median} with the 68, 90, and
100 Kaplan-Meier percentile envelopes (see Feigelson \& Nelson 1985,
Isobe, Feigelson, \& Nelson 1986) to allow comparison with other
samples. The median SED shows a large dispersion from the mean both in
the IR and optical/UV.

\section{The Infrared Spectral Energy Distributions.}

\subsection{Characteristics}

A closer look at the SEDs of the HEAO sample (Fig.~\ref
{fig:SEDs}) reveals that 13 out of 21 objects show either a
minimum at 1\micron\ (if the big blue bump is present) or a downturn
at 1\micron\ from the mid-IR (if the big blue bump is highly
reddened), consistent with a drop in opacity at the dust sublimation 
temperature $\sim$2000K (Sanders et al. 1989).  The remainder show
either no minimum/downturn (5 objects) or do not have enough data
points around 1\micron\ to detect the dip (3 objects). Three out of five
objects with no minimum/downturn at 1\micron, show steep ($\ge 2.5$)
far-IR cutoff slopes (MR~2251$-$178 has $\alpha_{cut}$=3.7, while
MCG$-$2$-$58$-$22 and E~1821+643 have $\alpha_{cut}$=2.5) indicating thermal
dust emission (see Introduction; far-IR cutoff slopes are presented in
Table~7). The remining 18 AGN have slopes flatter than
$\alpha_{cut}$=2.5, which can be interpreted as either dust emission
in which the lowest temperature dust emits at wavelengths longer than
our coverage (i.e. \lax~40\,K) or as inhomogeneous, non-thermal
synchrotron emission. While we cannot rule out non-thermal emission,
we assume the IR is dominated by dust emission in our ensuing
discussion.

The IR SEDs cover a broad wavelength range indicating a wide range of
temperatures in a thermal emission scenario. In 
Table~7 we list the maximum and minimum temperature of dust required
to match the continuum observed, estimated for each object by
comparing the observed data to grey-body curves at various
temperatures. An example is shown in Figure~\ref{temp:fg}. The lowest
indicated temperature is an upper limit when no long-wavelength
turnover is observed in the ISO band. The number of data points is
generally too few for formal fits to provide any useful constraints on
the dust temperature but these rough values, which are similar to
those reported in earlier studies, is too broad to be explained by
optically thick dusty torus models and is generally attributed to a
combination of AGN-heated torus emission and cooler emission from
starburst-heated dust (Rowan-Robinson 1995).

\subsection{Comparison with samples selected in other wavebands}

As discussed above, X-ray selection is expected to find a somewhat
different subset of the AGN population with less bias against sources
which include intrinsic l.o.s. absorption. In this section we compare
the SEDs of the HEAO sample with those of samples selected at optical,
near-IR and radio wavelengths in order to quantify this difference and
understand the subset of the population which may have been missed
from those samples.

We first compare the SEDs of our hard-X-ray selected AGN sample with
those presented by Elvis et al. (1994; hereafter E94). The E94 sample
consists of bright soft-X-ray objects observed by the {\it Einstein
Observatory} satellite, with sufficient counts to measure the
soft-X-ray slope. These objects were also chosen to be optically
bright enough to be observed by IUE. The sample is low redshift,
heterogeneous, half radio and half optically selected, and biased
towards AGN with strong soft-X-ray to optical flux ratios.

The HEAO and E94 samples have similar redshift ranges ($0 < z <0.35
$). We found (comparing our Table~5,6 and their Table~15,16) that the
range of IR and optical luminosities is similar in both samples, while
the UV, soft and hard-X-ray luminosities of the HEAO sample extend to
lower values (see also Figure~\ref{fig:L_z}).  We compared the
distributions of various IR, optical, UV octave luminosity ratios
(opt/near-IR, UV/near-IR, opt/UV, and IR/IR) between the two samples
and found that only the following luminosity ratios:
L(0.2-0.4$\mu$m)/L(0.4-0.8$\mu$m) and
L(0.2-0.4$\mu$m)/L(0.8-1.6$\mu$m) differ (see
Figure~\ref{fig:histo}). The Kolmogorov-Smirnov test shows a $<1$\%
probability that the HEAO and E94 samples have the same distributions
for these luminosity ratios. The HEAO sample extends to lower
L(0.2-0.4$\mu$m)/L(0.4-0.8$\mu$m) and
L(0.2-0.4$\mu$m)/L(0.8-1.6$\mu$m) ratios, meaning that it includes
objects with redder optical/UV continua.  This finding is also
confirmed when the median energy distribution of the HEAO sample is
compared with that of E94 (see our Fig.~\ref{fig:median} and their
Fig.~11a).  The discrepancy is strongest at the bluest wavebands
suggesting that dust absorption may be the cause.  In this case, the
range of L(0.2-0.4$\mu$m)/L(0.4-0.8$\mu$m) and
L(0.2-0.4$\mu$m)/L(0.8-1.6$\mu$m) ratios indicates that $\sim 30-40$\%
of the sample are reddened at levels of $E(B-V)\ge 1.2$ and may have
been missed by other surveys.

To understand the relation between the X-ray selected sources and
those from other samples we compare (Fig.~\ref{fig:B_K_mags}) the B-K
colors of the HEAO subset analyzed here (shaded area) with the full
HEAO AGN sample (dotted-line-shaded area), the radio selected quasars
from Webster et al. (1995; solid line), optically selected PG QSOs
from Neugebauer et al. (1987; dash-dot line) and the Chandra observed
subset of the 2~micron All Sky Survey (2MASS) AGN from Wilkes et
al. (2002).  The radio and hard X-ray selected samples show the widest
range of colors, though the radio selected sources have a larger
proportion of blue sources (one-tailed K-S test gave a 99.5\%
significance; mean B-K is: 3.34 and 4.03, while median B-K is: 3.1 and
3.99 for the radio and full HEAO sample respectively) similar to the
very narrow range of the optically selected PG sources. The PG quasars
have the bluest distribution: a one tailed K-S test gave $\gg$99.9\%
probability that the PG distribution is bluer than the radio, HEAO and
2MASS distributions respectively, which is also evident when the B-K
means are compared: 0.74(PG) with 3.34(radio), 4.03(full HEAO) , 5.60
(2MASS) and B-K medians: 0.72(PG) with 3.1(radio), 3.99(full HEAO),
5.48(2MASS). The red 2MASS AGN, selected by their red near-IR colors,
largely overlap the red end of the radio and hard X-ray selected
samples, implying that both these selection techniques include
similar, red AGN.

Although we cannot determine the B-K distribution of the intrinsic AGN
population from this comparison, Fig.~\ref{fig:B_K_mags} clearly
demonstrates the large discrepancies between samples selected in
different wavebands (a two-tailed K-S test showed that the
distributions of the PG, HEAO, 2MASS and Webster et al. radio samples
have different distributions at the $\gg$99.5\%). It is clear that
radio and hard X-ray samples cover a much wider range of optical
colors than optically selected samples. It is also notable that the
peak in the distribution of B-K colors for the hard X-ray selected
sources is significantly redder than that of the radio selected sample
(one-tailed K-S test gave $\gg$99.5\% significance). This indicates
either a lack of blue AGN in the hard X-ray sample, which seems
unlikely given that these are the easiest AGN to find and classify, or
a lack of red AGN in the radio sample.  Such a lack could be due to
differing classification schemes for radio and X-ray selected sources
preferentially excluding more of the reddest sources in the radio
sample {\it e.g.} due to the lack of broad optical emission lines.

\section{Implications for the structure of Active Galactic Nuclei.}

\subsection{A comparison of dust (optical) versus gas (X-ray) absorption
column densities.}

If the redder B-K colors indicate more dust obscuration, as suggested
by Webster et al. (1995), then the above findings suggest that the
HEAO sample includes a larger proportion of dust obscured AGN than
either optically or radio selected samples which miss perhaps as many
as $\sim 30-40$\% of the population.  The availability of both optical
and X-ray data for many of these sources allows us to investigate the
properties and structure of both the dust and the gas in the obscuring
material.  It is well-established that AGN show a smaller amount of
optical dust absorption than indicated by the X-ray (gas) absorption
if a `normal' gas-to-dust ratio is assumed (Maiolino et al. 2001,
Risaliti et al. 2000). The ratio varies from factors of a few to 2-3
orders of magnitude. A number of explanations have been suggested such
as abnormal gas-to-dust ratios, largely dust-free gas, dust in the
form of large grains whose opacity is not a strong function of
wavelength and differing lines-of-sight and corresponding absorption
column densities to the optical and X-ray emission regions.

A subset of the sources in the current sample have sufficient data to
make a comparison. Those objects with known intrinsic X-ray
absorption, are: 3A~05574$-$383 (Sy1 with
$N_H=7\times10^{21}$~cm$^{-2}$), IC~4329A (Sy1.2 with
$N_H=1.5\times10^{21}$~cm$^{-2}$), and H1834$-$653 (Sy2 with
$N_H=13.5\times10^{22}$~cm$^{-2}$). Other AGN show either no evidence
of intrinsic X-ray absorption or no information is available for them
(see Table~\ref{tab:ebv_star}).  As dust is generally associated with
gas, objects with intrinsic $N_H$ are expected to show redder
optical/UV spectra. Indeed those objects with intrinsic $N_H$ show the
lowest L(0.2-0.4$\mu$m)/L(0.4-0.8$\mu$m) and
L(0.2-0.4$\mu$m)/L(0.8-1.6$\mu$m) ratios (see Fig.~\ref{fig:histo}
where the high intrinsic $N_H$ objects are represented by hatched
regions). To estimate the amount of intrinsic dust absorption, we took
the observed SEDs (previously corrected for Galactic extinction) and
dereddened them by a range of $E(B-V)$ values starting from 0.1 to 1.5
in steps of 0.1. The smallest $E(B-V)$ which produced an SED with a
big blue bump showing no downturns in the optical/UV was taken as a
lower limit estimate of the intrinsic absorption for that object. For
IC~4329A and H1834$-$653 the $E(B-V)$ was estimated to be $\ge 1.2$
and $\ge 1.1$ respectively, which translates to $N_H = 6\times10^{21}$
and $5.5\times10^{21}$~cm$^{-2}$, assuming the standard conversion of
$N(HI)/E(B-V)=5\times10^{21}$cm$^{-2}$ for Galactic dust/gas ratio
(Diplas \& Savage 1994; 3A~05574$-$383 has too few opt/UV data points
to estimate the big blue bump shape). A comparison of absorption
estimated from the opt/UV (above) and X-rays
(Table~\ref{tab:ebv_star}) shows a $\sim$4 times higher and a 30 times
lower dust/gas ratio for IC~4329A and H1834$-$653 respectively. A
higher/lower dust to gas ratio can be explained by either
smaller/larger dust grains or the presence of an ionized absorber (see
below) in the higher dust to gas ratio objects.

In our sample we also find objects which are extremely red in the
optical/UV wavelength range, with non-existent big blue bumps (where
L(0.2-0.4$\mu$m)/L(0.4-0.8$\mu$m) and
L(0.2-0.4$\mu$m)/L(0.8-1.6$\mu$m) is less than $-$0.5) but normal
X-ray softness ratio ($L(0.5-2keV)/L(2-10keV) \ge 0$) and steep
($\Gamma > 1$) soft-X-ray slopes indicating no/little intrinsic X-ray
absorption from neutral gas. These include: Mkn~1152 (Sy1.5) and
MCG$-$6$-$30$-$15 (Sy1.2). The lower limit for $E(B-V)$ estimated from
the downturn of the big blue bump in the optical/UV range is 1.2 and
1.6, for Mkn~1152 and MCG$-$6$-$30$-$15 respectively. This translates
to HI column densities of $6\times10^{21}$~cm$^{-2}$ and
$8\times10^{21}$~cm$^{-2}$ respectively (assuming a Galactic dust/gas
ratio), which should result in detectable soft-X-ray absorption in the
spectra of these objects. MCG$-$6$-$30$-$15 is known to have a dusty
warm absorber (Reynolds 1997), which can explain the lack of
soft-X-ray absorption via neutral gas combined with the reddening in
opt/UV via dust. Since Mkn~1152 shows similar opt/UV/X-ray SED
properties to MCG$-$6$-$30$-$15 it may be a candidate for a dusty warm
absorber. However detailed X-ray spectral analysis is needed to
confirm this and rule out smaller than Galactic dust grains as another
option.
In addition IC~4329A, noted above to show X-ray absorption, but with
dust/gas ratio a few times higher than the Galactic value, can also be
explained if some of the absorbing gas is ionized.  The X-ray spectrum
of this object, observed by BeppoSAX (Perola et al. 1999), shows
comparable column densities of both cold and warm absorption (the
latter indicated by the $\sim$0.7\,keV feature corresponding to the
O\,VI and O\,VII blend) and so is consistent with our interpretation.


To obtain a more general view of the sample we plot
(Fig.~\ref{fig:04_02_hardness}) the X-ray softness ratio i.e. the
L(0.1-2keV)/L(2-10keV) as a function of the big blue bump shape
indicated by the following UV/optical/near-IR luminosity ratios:
L(0.2-0.4$\mu$m)/L(0.4-0.8$\mu$m) and
L(0.2-0.4$\mu$m)/L(0.8-1.6$\mu$m).  AGN with high ($>$0)
L(0.2-0.4$\mu$m)/L(0.4-0.8$\mu$m) and
L(0.2-0.4$\mu$m)/L(0.8-1.6$\mu$m) ratios, which are also comparable to
the luminosity ratios of the soft-X-ray selected sample of E94 (see
Figure{~\ref{fig:histo}}), are those with pronounced, and unabsorbed
big blue bumps, and include mostly Sy1s (open circles in Fig.~8 with
the exception of Mkn~509 a Sy1.2, and MCG$-$2$-$58$-$22 a Sy1.5). AGN
with ratios $<0$, and smaller than E94, are mostly intermediate type
(Sy1.2, Sy1.5) and type 2 Seyferts (indicated in
Figure~\ref{fig:04_02_hardness} by filled circles). This finding is
broadly consistent with unified models of AGN in which Sy1s are viewed
face on to a dusty torus-like structure, and hence their continua are
expected to be less absorbed and bluer than the continua of the
intermediate type and type 2 Seyferts, which are viewed at larger
(more edge-on) inclination angles, and are more absorbed.  Objects
with low X-ray softness ratios ($< -0.5$) in
Fig.~\ref{fig:04_02_hardness}, indicative of neutral gas absorption,
include: H1834-653 - a heavily absorbed Seyfert 2, 3A~05574$-$383 - a
Seyfert 1 with considerable $N_H$ (see Table~\ref{tab:ebv_star}),
H1537+339 (Sy1) and H1419+480 (Sy1) for which no indication of
intrinsic $N_H$ has been found in the literature. Unfortunately for
almost all of these objects (with the exception of the H1834-653) the
SEDs in the 0.2$-$0.4$\mu$m and UV wavelength range lack data points,
so the values of L(0.2-0.4$\mu$m)/L(0.4-0.8$\mu$m) and
L(0.2-0.4$\mu$m)/L(0.8-1.6$\mu$m) ratios in
Fig.~\ref{fig:04_02_hardness} are uncertain. We can however predict
the direction (indicated by horizontal arrows) in which the SED would
move if the UV points were available based on the blue bump shape in
the optical. Objects which lack soft-X-ray data, i.e. those which have
uncertain softness ratios, are indicated by vertical arrows. We also
plot in Fig.~\ref{fig:04_02_hardness} the dependence of UV/optical
colors and X-ray softness ratio on $N_H$. It is apparent that the
UV/optical colors can be explained by a lower $N_H$ range, than is
needed to explain the X-ray softness range.

\subsection{Diagnostics from the IR continuum}

The relative contribution of emission in the warm (mid-IR: $\sim 25
\mu$m) and cool (far-IR: $\sim 60-100 \mu$m) IR potentially provides
an important diagnostic of the energy sources responsible. The
relative strength of the mid- to far-IR emission correlates with that
of an active nucleus (de Grijp et al. 1985, Heisler \& De Robertis
1999). In the popular, two-component thermal dust model for AGN IR
emission (Rowan-Robinson \& Crawford 1989, Rowan-Robinson 1995), the
AGN heats dust close to the nucleus generating mid-IR emission in
addition to the far-IR emission from a ubiquitous starburst 
component.  As we have seen, a popular model for the AGN IR component
is re-emission from an optically thick dusty torus (Pier \& Krolik
1992, Granato \& Danese 1994) which unifise type 1 and type 2 AGN via
obscuration of the central AGN by the edge-on torus in the latter.
The torus optical depths are sufficiently high (\nh $\sim 10^{24}$
cm$^{-2}$) that the SED of the AGN IR component is a strong function
of viewing angle or, equivalently, the column density of obscuring
material along our l.o.s.  through the material.  It is likely that both
AGN luminosity and the amount of l.o.s. obscuration contribute to the
mid-IR SED shape, combined with the relative strength of the host
galaxy emission at the lower luminosity end (Lumsden et al. 2001,
Alexander 2001).

We determine the luminosities of the HEAO sample in the warm
3-60\micron\ band and cold 60-200\micron\ bands
(Table~\ref{tab:integrIR}) and compare this ratio with those for the
quasar sample of Polletta et al. (2000), which includes both
radio-loud and radio-quiet quasars with SEDs of similar quality to
ours.  The resulting ratio of warm to cool IR emission is plotted in
Figure~\ref{fig:Poletta} as a function of the total IR luminosity for
both samples. The distribution suggests that the HEAO sample lacks
objects with large L(3-60$\mu$m)/L(60-200$\mu$m) ratios i.e. those
with the largest AGN contribution, although this difference is only
marginally significant for the samples as a whole (the K-S test gave a
5\% probability that the two samples have the same
distribution). Since the non-thermal radio emission has been
subtracted from the IR spectra of the Polletta et al. (2000) quasars,
and objects with larger L(3-60$\mu$m)/L(60-200$\mu$m) ratios are
either radio-quiet or radio intermediate, any difference cannot be due
to the contribution of radio-linked, non-thermal emission in their
radio-loud quasars. Those objects with the largest AGN contribution
(L(3-60$\mu$m)/L(60-200$\mu$m)$>1.7$) are those with the highest total
IR luminosity supporting a scenario where the far-IR emission is
associated with a starburst in the host galaxy which is more dominant
for the lower luminosity and/or intermediate type sources of which the
HEAO sample includes a larger proportion.

In Table~\ref{tab:integrIR} we also present a widely-used alternative
measure of warm to cool IR emission, the ratio between the 25\micron\
and 60\micron\ flux, and compare (Figure~\ref{fig:alpha_25_60}) the
HEAO sample with the soft-X-ray sample of E94, Seyfert galaxies
(including type 1 and 2) from de Grijp et al. (1985), and Low et
al. (1988), and the Palomar-Green QSOs from Sanders et al. (1989). In
this figure we also mark the ratio characteristic for normal spiral
galaxies in the Virgo cluster from Soifer et al. (1987). The HEAO
25/60 ratios extend towards warmer colors when compared to Seyfert
galaxies, but are consistent with the IR colors of QSOs and the E94
sample. Although one object (3A~0557$-$385) in the HEAO sample has a
larger 25/60 ratio than the E94 sample, and E94 AGN extend towards
smaller ratios, the distribution of the 25/60 ratios is statistically
indistinguishable in HEAO and E94 samples (K-S test gives a 84\%
probability that the samples are drawn from the same population).

If obscuration is a significant contributor to the 25/60 flux ratio,
then we would expect some difference between these distributions for
the HEAO and E94 samples. We showed in section~3.2 that the HEAO
sample includes more optically reddened AGN than the E94 sample, with
half of the HEAO AGN showing lower UV/optical ratios
(i.e. L(0.2-0.4$\mu$m)/L(0.4-0.8$\mu$m) and
L(0.2-0.4$\mu$m)/L(0.8-1.6$\mu$m)) than the E94 AGN.  We plot in
Fig.~\ref{fig:25_60tooptUV} the relation between 25/60 ratio and the
UV/optical ratio: L(0.2-0.4$\mu$m)/L(0.4-0.8$\mu$m) and
L(0.2-0.4$\mu$m)/L(0.8-1.6$\mu$m).  The HEAO AGN with
L(0.2-0.4$\mu$m)/L(0.4-0.8$\mu$m)$<0$ are mostly intermediate and type
2 Seyferts. No correlation between the 25/60 ratio and the dust
indicator from the UV/optical SED is present, the more reddened
Seyferts (with L(0.2-0.4$\mu$m)/L(0.4-0.8$\mu$m)$<0$) do not show
larger 25/60 ratios than the blue
(i.e. L(0.2-0.4$\mu$m)/L(0.4-0.8$\mu$m)$>0$) Seyferts (K-S test showed
a less than 30\% probability that the two distributions are
different). The combination of redder optical colors and lack of a
difference in the mid-IR indicates \nh $\sim 10^{22}$~cm$^{-2}$.  This is
inconsistent with torus models implying an obscuring structure which
is less optically thick, at least along the lines-of-sight represented
here.

In Fig.~\ref{fig:12_60to25_60} we present the relation between 12/60
and 25/60 ratios and represent the HEAO Sy1s as open circles, HEAO
intermediate type Sys as filled circles, Sy2 as filled square, and the
E94 AGN as crosses. We also include from Heisler, Lumsden and Bailey
(1997) the Seyfert~2s with a hidden broad line region (HBLR) i.e.
Seyfert~2 galaxies, that show broad emission lines in polarized light,
and Seyfert~2s without a hidden broad line region (non-HBLR)
i.e. Seyfert~2 galaxies which do not show broad emission lines in
polarized light.  This correlation between the 12/60 and 25/60 ratios
in HBLR and non-HBLR Seyfert~2 galaxies was initially interpreted by
Heisler et al. (1997) within the context of unified schemes, such that
Seyfert~2 galaxies with no HBLR are viewed edge-on to the torus, which
obscures the mid-IR emission and hides the scattering medium producing
polarized broad emission lines, while the Sy2s with HBLR are viewed
from angles skimming the dusty obscuring structure, large enough to
hide the broad line region, and small enough to see the warm
25\micron\ dust emission and the scattering medium (see also Tran 2003
who finds hotter circumnuclear dust temperatures and mid-IR spectra
more characteristic of Sy1s in HBLR Sy2s compared to non-HBLR
Sy2s). This simple picture was later modified to include the
significant effects of the host galaxy in the non-HBLR Sy2s which tend
to have lower overall luminosity (Alexander 2001, Lumsden et
al. 2001). The HEAO and E94 AGN extend the non-HBLR and HBLR Seyfert 2
correlation towards warmer 25/60 and 12/60 ratios roughly along a
reddening curve for Galactic dust.  This general trend again points
towards a modified unified scheme including an obscuring structure
which is less optically thick than a torus and whose column density
decreases more smoothly towards more face-on viewing angles.

In Fig.~\ref{fig:12_60to25_60} there is a region, at colors suggesting
\nh$\sim10^{23}$ according to the Galactic reddening curve shown,
where there is overlap between the reddest type 1 AGN from both E94
and the HEAO sample, intermediate types from the HEAO sample and the
HBLR Sy2s from Heisler et al. (1997). In this scenario, in which the
viewing angles are similar for these objects, such a mix could be
achieved via differing lines-of-sight to the BLR and the various
continuum emitting regions. An alternative is to invoke a clumpy
distribution of dust, where the Sy1-Sy1.5 in the HEAO sample are
viewed through holes, while the HBLR Sy2s are totally obscured.

Looking at a few of the individual cases could prove instructive here.
Along with two HBLR Sy2s, the spectroscopically classied Sy2 H1834-653
has a warm 25/60 ratio of $-0.03$, more consistent with the colors of
Sy1s.  It is one of the lowest luminosity sources in our sample with a
high host galaxy/AGN ratio (see Tables~5 \& 3) so the AGN-like
appearance of its mid-IR colors cannot be explained by an unusually
dominant AGN in the IR. However the amount of reddening needed to
explain the red opt/UV SED is $\sim$ E(B-V)=1.1 (section~4.1), which
corresponds to a negligible extinction of 0.05~mag at 25\micron\ and
0.007~mag at 60\micron, is likely sufficient to allow the galaxy light
to dominate the optical/UV emission and thus to lose the BLR.
Alternatively the obscuring material could lie in the host galaxy, as
in IC~4329A, a Sy1.2 with a similar 25/60 ratio (i.e. similar viewing
angle) to H1834-653 and for which the AGN is viewed through an edge-on
dust lane in its host galaxy (Wolstencroft et al. 1995). By contrast
there are two broad-lined AGN (3C48 and PG1613+658) in the E94 sample
which have mid-IR colors similar to Sy2s. Again the relative AGN and
host galaxy luminosities cannot explain the galaxy-like IR colors in
these sources, both have optical-mid-IR luminosities in the AGN range
($\sim 10^{45}$ erg s$^{-1}$) and small host galaxy contributions.
3C48 is a steep spectrum compact radio source, one popular model for
which involves their confinement by an inhomogeneous, dense medium in
the host galaxy (van Breugel et al. 1984), potentially explaining the
reddening of the far-IR continuum, though the clear view of the BLR
remains a puzzle. For PG1613+658 there is no obvious explanation
beyond invoking an unusually strong far-IR starburst component as
suggested by the detection of CO in this source (Evans et al. 2001).

Summarizing the last two sections, a simple unified model can not
explain the behavior of the HEAO SEDs. A relation between Seyfert type
and opt/UV reddening is present in the sense that bluer AGN are mostly
Seyfert~1s, and redder AGN are mostly intermediate or type 2
Seyferts. This relation does not extend to the infrared in a way
consistent with standard optically thick torus models as the 25/60
ratios are similar in Seyferts with differing type and optical/UV
reddening (except for non-HBLR Seyferts). However they can be
explained by a lower optical depth, disk-like structure whose
obscuring column density is a function of viewing angle. If we further
include the effects of differing ratios of host galaxy to AGN
luminosity, most of the observed properties can be explained.

\subsection{Models}  

The lower optical depths (\nh $\sim 10^{22-23}$ cm$^{-2}$) for the
obscuring material favored by our results suggest that the AGN may be
able to heat a larger amount of dust to a larger range of temperatures
than in the optically thick torus models. This raises the question of
whether/not a cool, starburst component is required to explain the
full IR continuum of AGN.  To investigate this we use the Monte Carlo
radiative equilibrium technique of Bjorkman \& Wood (2001), modified
to work in the parameter regime of AGN, to compute model spectral
energy distributions (SEDs). In our models we assume that an isotropic
point source illuminates an axisymmetric dusty disk-like structure.
Incident spectra for the central engine in AGN SED models are often
power laws or black-bodies (e.g., Granato \& Danese 1994).  Instead,
in our models, we use the median SED of a normal, broad-line AGN from
Elvis et al. (1994) as the incident continuum.

The dust model is that presented by Kim, Martin, \& Hendry (1994) and
provides a good match to the observed extinction and scattering
properties of dust in the Milky Way's ISM.  Our SED models suffer from
the same problem as other axisymmetric models in that they predict
silicate emission at 10$\mu$m which is not generally observed.  Recent
work by Nenkova, Ivezic, \& Elitzur (2002) suggests that this problem
may be alleviated with three dimensional multiple cloud models and
does not require different dust species or environments for Seyfert 1
and 2 models (e.g., Efstathiou \& Rowan-Robinson 1995).  There have been
many theoretical studies of the shapes of SEDs from different
axisymmetric dusty tori (e.g., Pier \& Krolik 1993; Efstathiou \&
Rowan-Robinson 1995; Granato \& Danese 1994) and more recently three
dimensional multiple dust cloud models (Nenkova, Ivezic, \& Elitzur
2002).  Our SED modeling, while using a different technique for the
numerical radiation transfer, adopts an axisymmetric structure
(hereafter called dusty disk) for the torus as in the papers cited 
above. We adopt the following density structure:
\begin{equation}
\rho=\rho_0 \left ({R_0/{\varpi}}\right )^{\alpha}
\exp{ -{1\over 2} [z/h(\varpi )]^2  }
\; ,
\end{equation}
In this equation we work in units of $R_0 = 1$~pc, $\rho_0$ is the
dusty disk density at $R_0$, $\varpi$ is the radial coordinate in the
disk mid-plane and the scale height increases with radius, $h=h_0\left
( {\varpi /{R_0}} \right )^\beta$. We adopt $\alpha = 0$ (radial
density independent of distance) and a slightly flaring scaleheight
with $\beta = 1.1$ and fix the radial extent of the disk at $R_{max} =
300$~pc.  The parameters that we varied to fit the three SEDs are the
incident luminosity $L_{AGN}$, scaleheight $h_0$, and total mass of
the dusty disk $M_{torus}$.  The other parameter in our modeling is
the disk inclination $i$.

We have modeled three AGN which span the range of SED types observed
in our sample.  PG~0804+761 is a typical SED, MR~2251$-$178 shows a
rather flat SED, and IR~13218+0552 is very red.  Our model fits to
these three AGN are shown in Figure~\ref{fig:kw1}.  Our goal in this
modeling is to reproduce the overall shape of the different SEDs
varying as few parameters as possible.  Our results are certainly not
unique (e.g., see Granato \& Danese 1994 for the effects of changing
model parameters), but they do allow us to estimate dusty disk
parameters for the three AGN: the dusty disk mass and shape
($M_{torus}$, $h_0$) is obtained from the level of the mid and far-IR
emission.  The change in SED as a function of inclination angle for
the best fit model for each AGN is shown in Figure~\ref{fig:kw2}. For
PG~0804+761 and MR~2251$-$178 pole-on views satisfactorily fit the
SED, while for IR~13218+0552, we require a sightline that obscures the
central engine at optical wavelengths.  In all cases the model SED can
reproduce the full AGN SED with no requirement for an additional
starburst component. The resulting model parameters are given in
Table~8. The temperature and density distribution of dust in the
obscuring disk-like structure are shown in Fig.~\ref{fig:kw3}.  As is
immediately evident the hottest dust lies at the surface and inner
edge of the disk while the coolest dust with highest density is
located in the plane of the disk.

\section{Summary}

We have studied the IR--X-ray properties of 21 HEAO1-A2 AGN. The
sample is hard-X-ray selected and hence relatively unbiased by the
effects of obscuration.  New ISO observations were compiled with
literature data to obtain the most comprehensive radio to hard-X-ray
SEDs, for comparison with the SEDs of radio, optically, and soft-X-ray
selected samples. We found that the HEAO AGN include objects with
redder opt/UV continua than the optically and soft-X-ray selected AGN,
with a B$-$K color distribution similar to radio and near-IR (2MASS)
selected AGN, although the peak of the distribution in the
radio-selected samples is shifted to bluer objects. We hence estimate
that $\sim30-40$\% of the redder AGN may be missed by the optical and
radio surveys.  The bluer objects in the HEAO sample are mostly
Seyfert~1 objects, while the redder are mostly intermediate and type 2
Seyferts, indicating that Sy1s have less obscured nuclei than the
Sy2. This trend does not extend, however, into the infrared where the
25/60 flux ratios do not differ much between Seyfert type and are
consistent with the ratios found in the unobscured optically selected
QSO and soft-X-ray selected AGN.  There is also no correlation between
the 25/60 flux ratio and the dust indicator from the opt/UV SEDs,
indicating column densities of the order of $N_H \sim
10^{22}$~cm$^{-2}$, much lower than predicted by the standard compact
($\sim$few parsec) torus model. Such a behavior can be explained if
the obscuring dusty disk-like structure has a lower optical depth and
extends towards larger radii ($\sim$~few hundred parsecs) than the
standard torus, with a possibly clumpy dust distribution.  We show
that such a model fully explains the large range of IR SED observed in
our HEAO sample, and since dust can be heated at larger distances,
there is no need to introduce an additional starburst component to
account for the far-IR emission. The range of dust temperatures is
high ranging from 20K to 1000K and the far-IR turnover occurs at
wavelengths longer than  200$\mu$m  in all but 3 objects.

\acknowledgements

We wish to thank an anonymous referee for comments that helped to
improve the manuscript. The authors gratefully acknowledge support
provided by NASA grant NAG5-8847 (ISO), and NAG5-6410 (LTSA).

\clearpage


\clearpage
\begin{figure}[t]
\epsscale{1.1}
\plotone{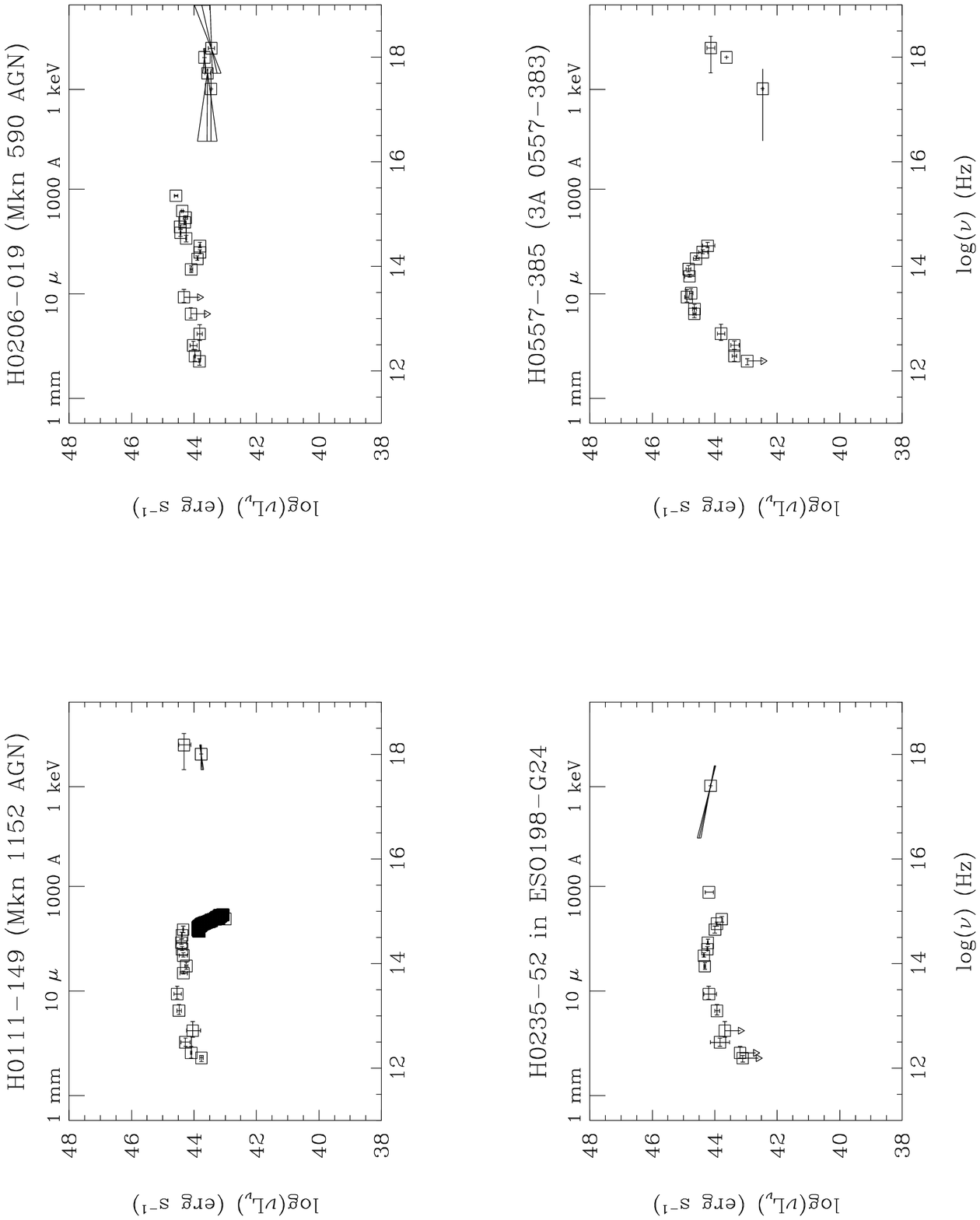}
\caption{Radio to X-ray spectral energy distributions (SEDs)
for AGN in our sample on a log$\nu$L$_{\nu}$ vs log$\nu$
scale.}
\label{fig:SEDs} 
\end{figure}

\begin{figure}[t]
\epsscale{1.1}
\plotone{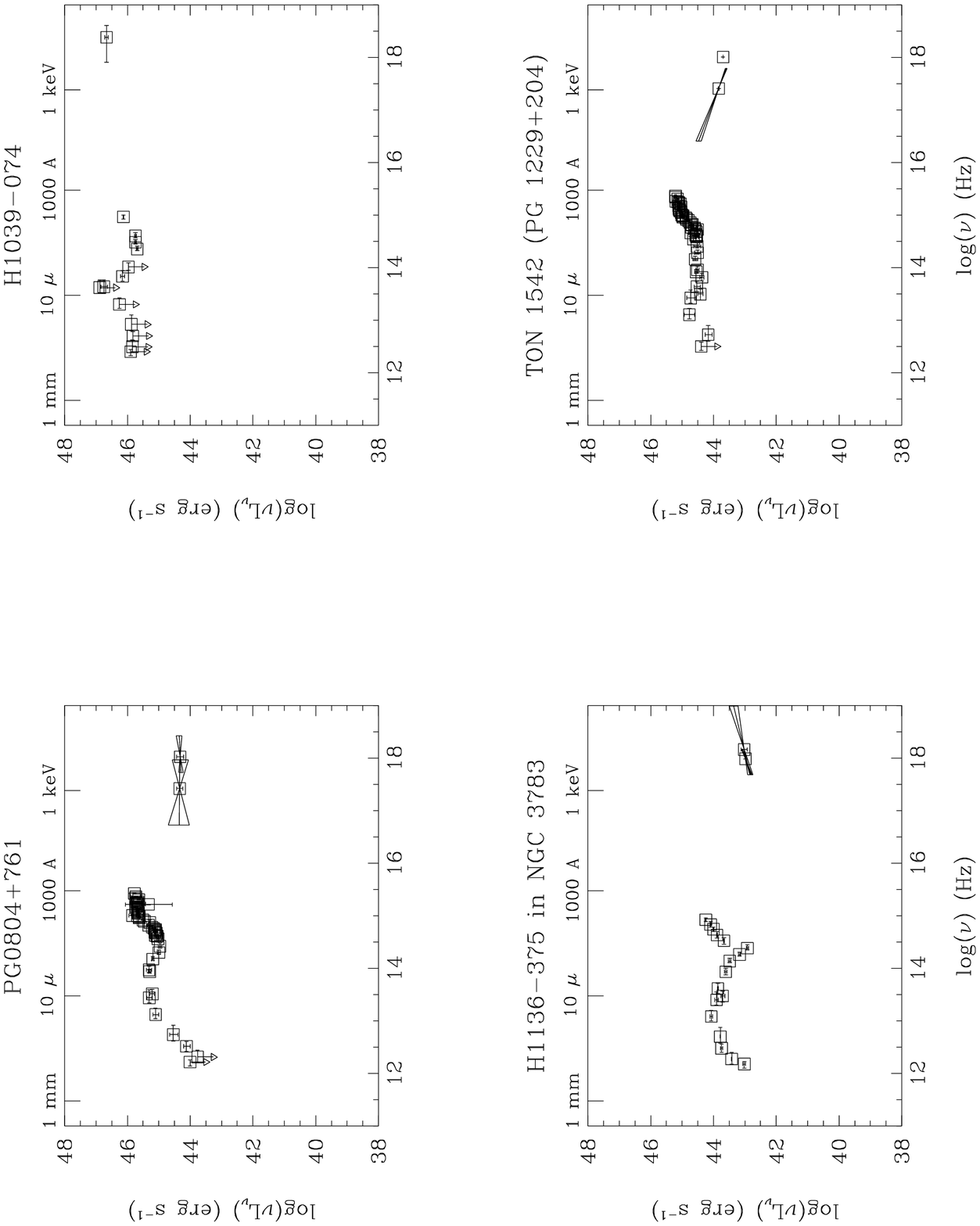}
\setcounter{figure}{0}
\caption{--continued}
\end{figure}

\begin{figure}[t]
\epsscale{1.1}
\plotone{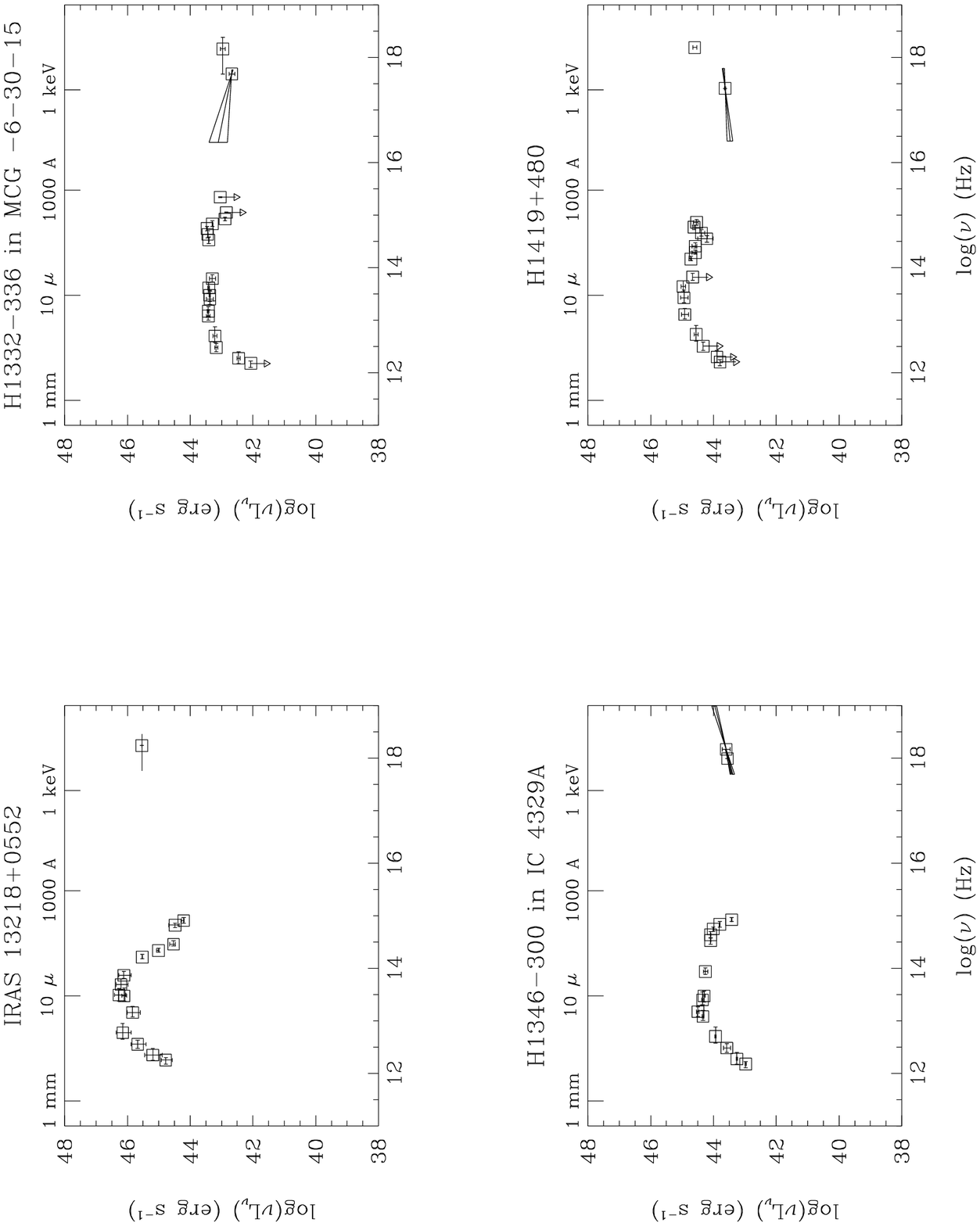}
\setcounter{figure}{0}
\caption{--continued}
\end{figure}

\begin{figure}[t]
\epsscale{1.1}
\plotone{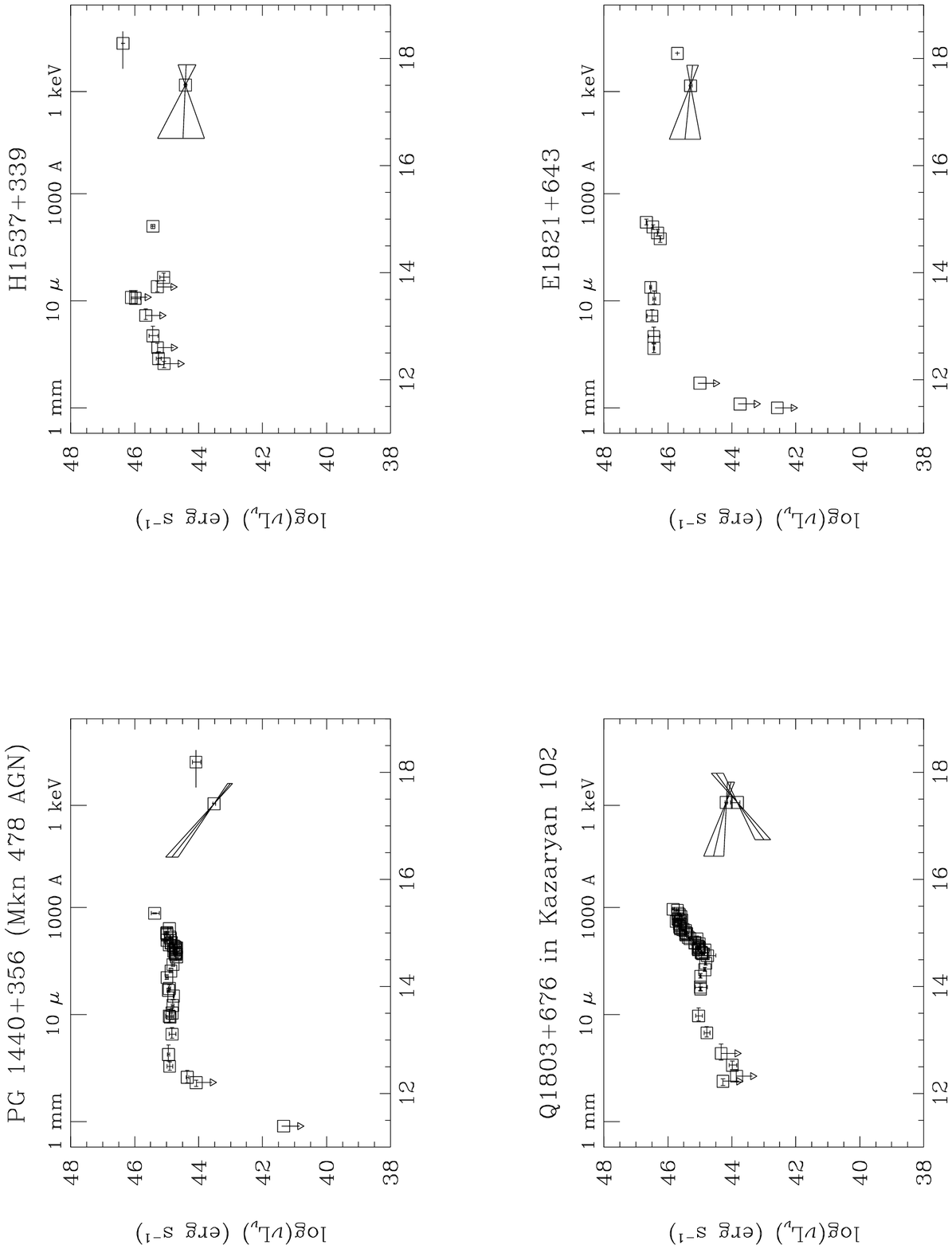}
\setcounter{figure}{0}
\caption{--continued}
\end{figure}

\begin{figure}[t]
\epsscale{1.1}
\plotone{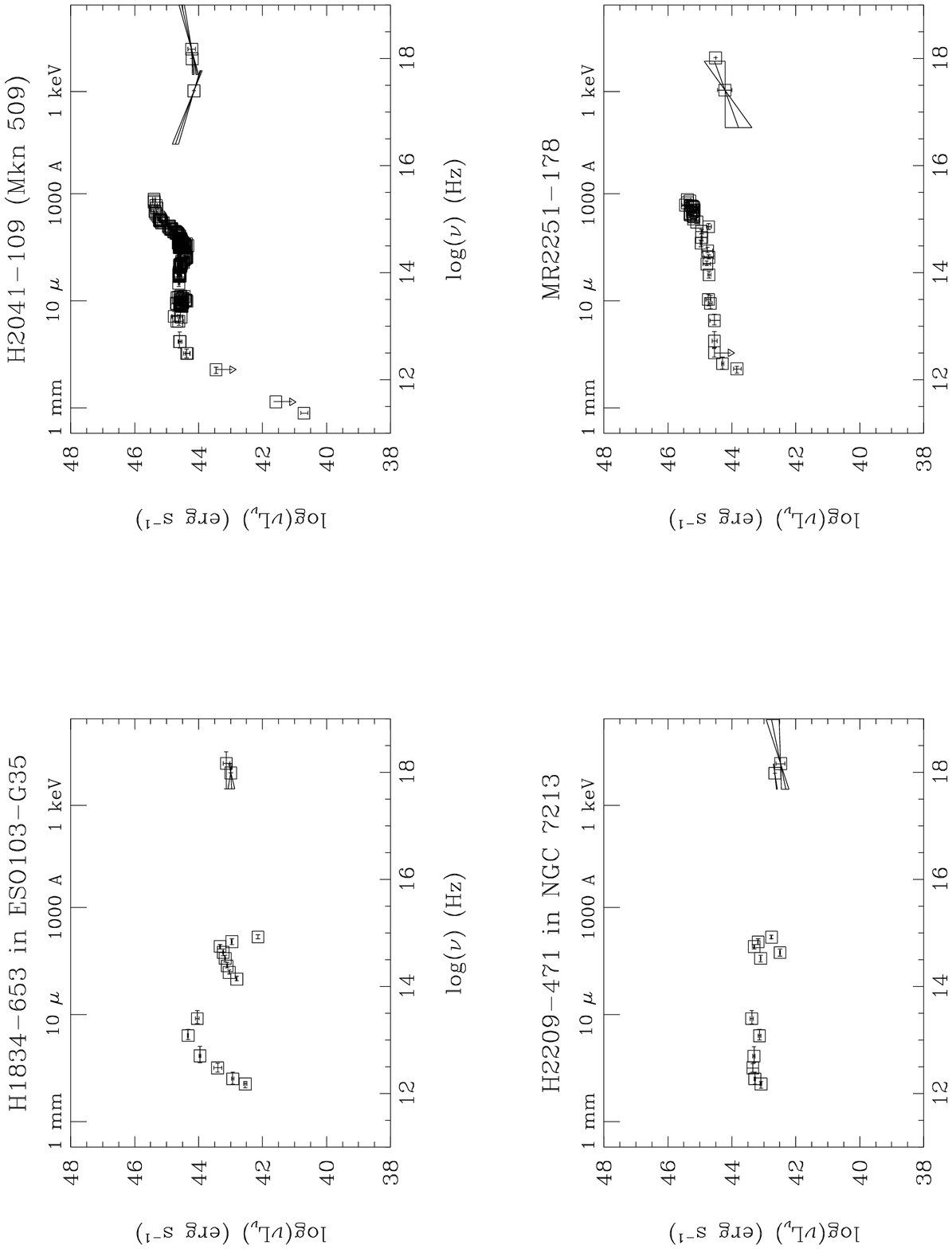}
\setcounter{figure}{0}
\caption{--continued}
\end{figure}

\begin{figure}[t]
\epsscale{1.1}
\plotone{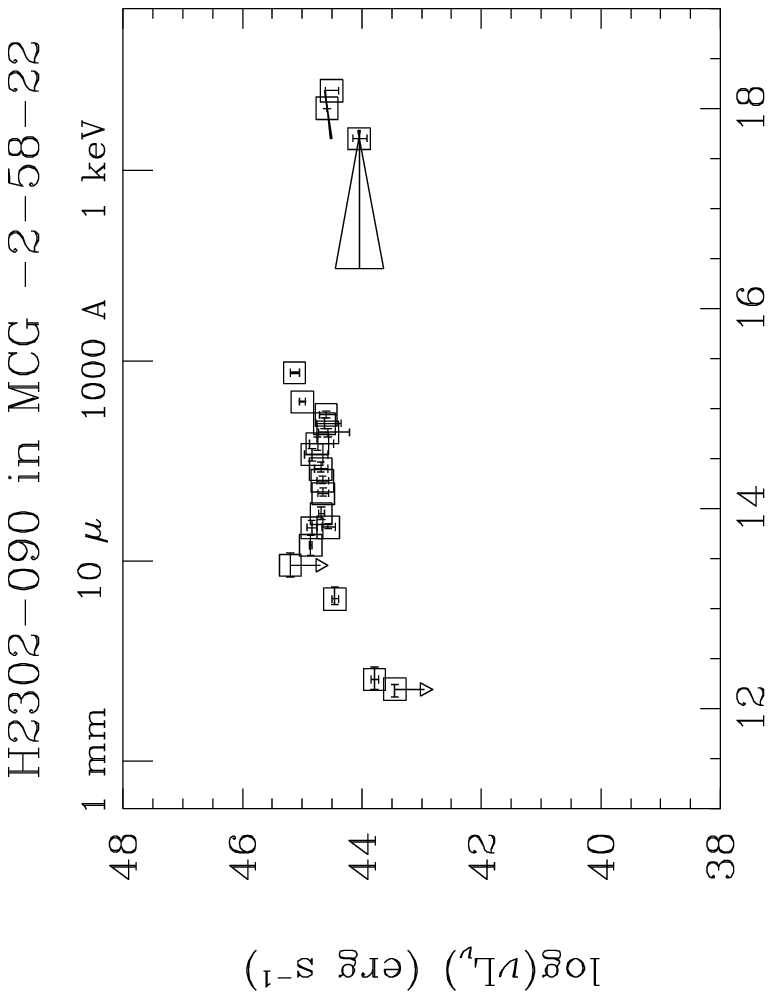}
\setcounter{figure}{0}
\caption{--continued}
\end{figure}

\epsscale{0.85}
\begin{figure}
\plotone{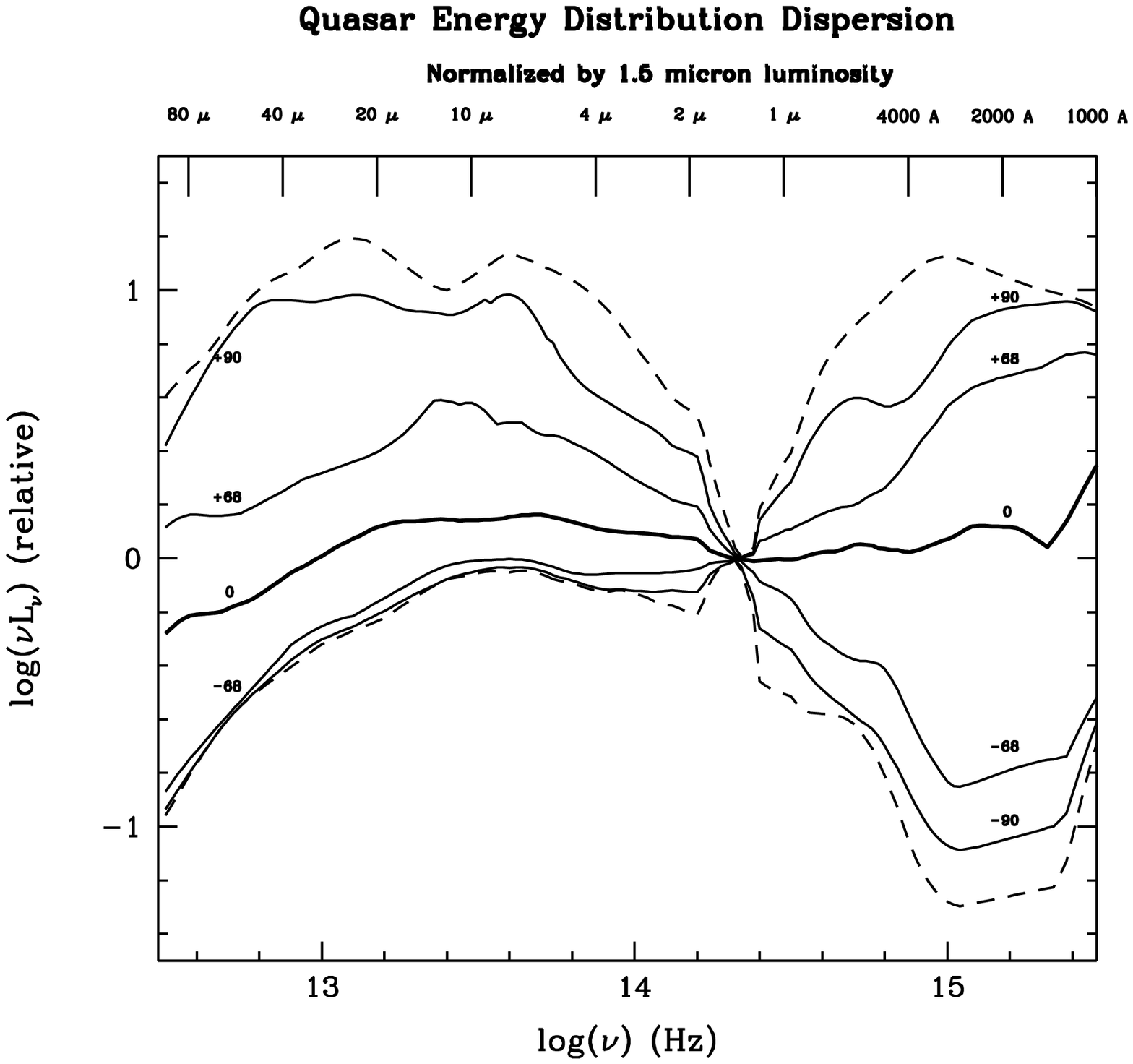}
\caption{a) Mean energy distributions for our sample normalized at
1.5~$\mu$m and the 68, 90, and 100 (dashed line) Kaplan-Meier
percentile envelopes.}
\label{fig:median} 
\end{figure}

\begin{figure}
\plotone{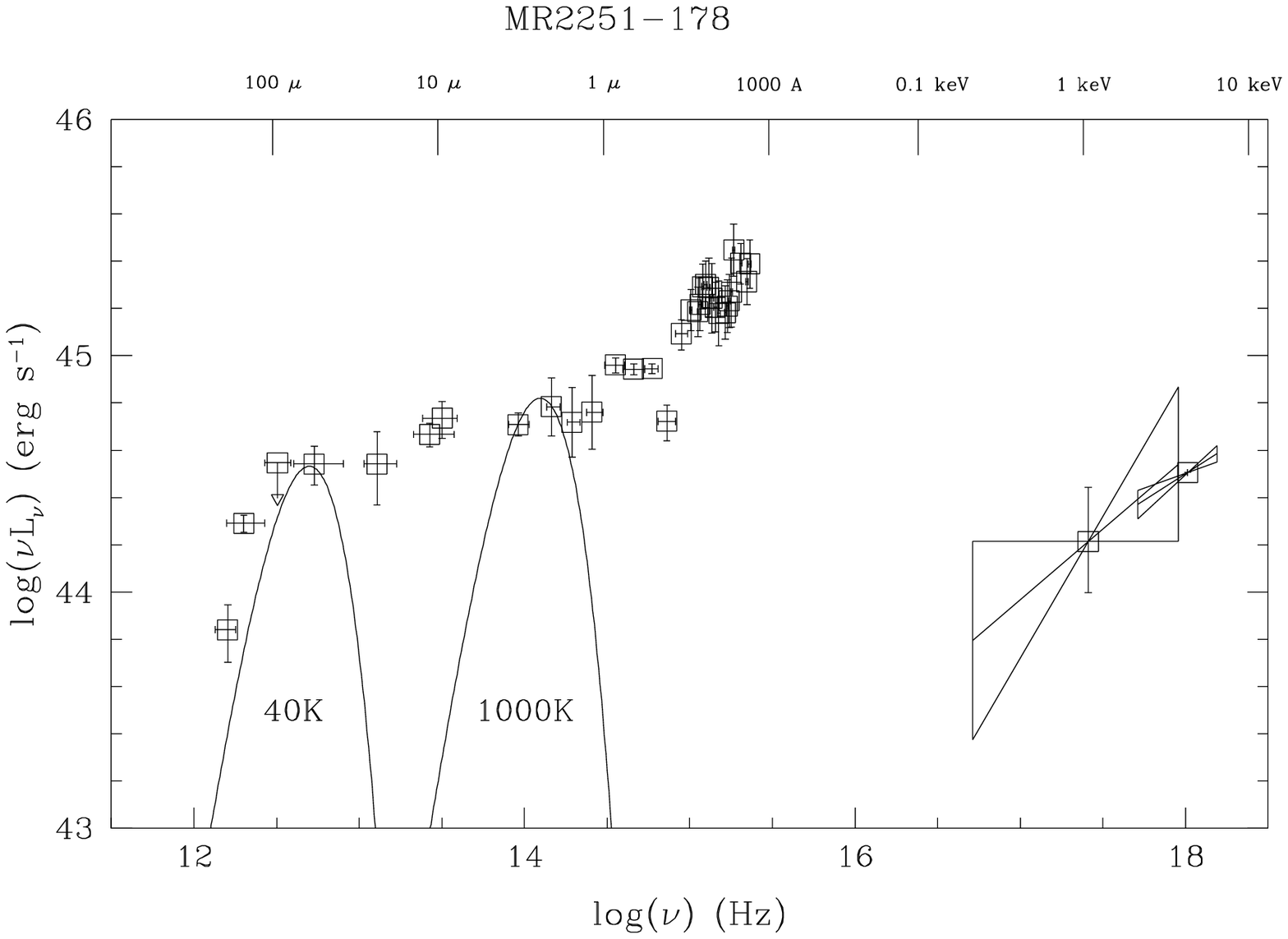}
\caption{The dust temperature range in MR~2251$-$178.}
\label{temp:fg}
\end{figure}

\begin{figure}
\plotone{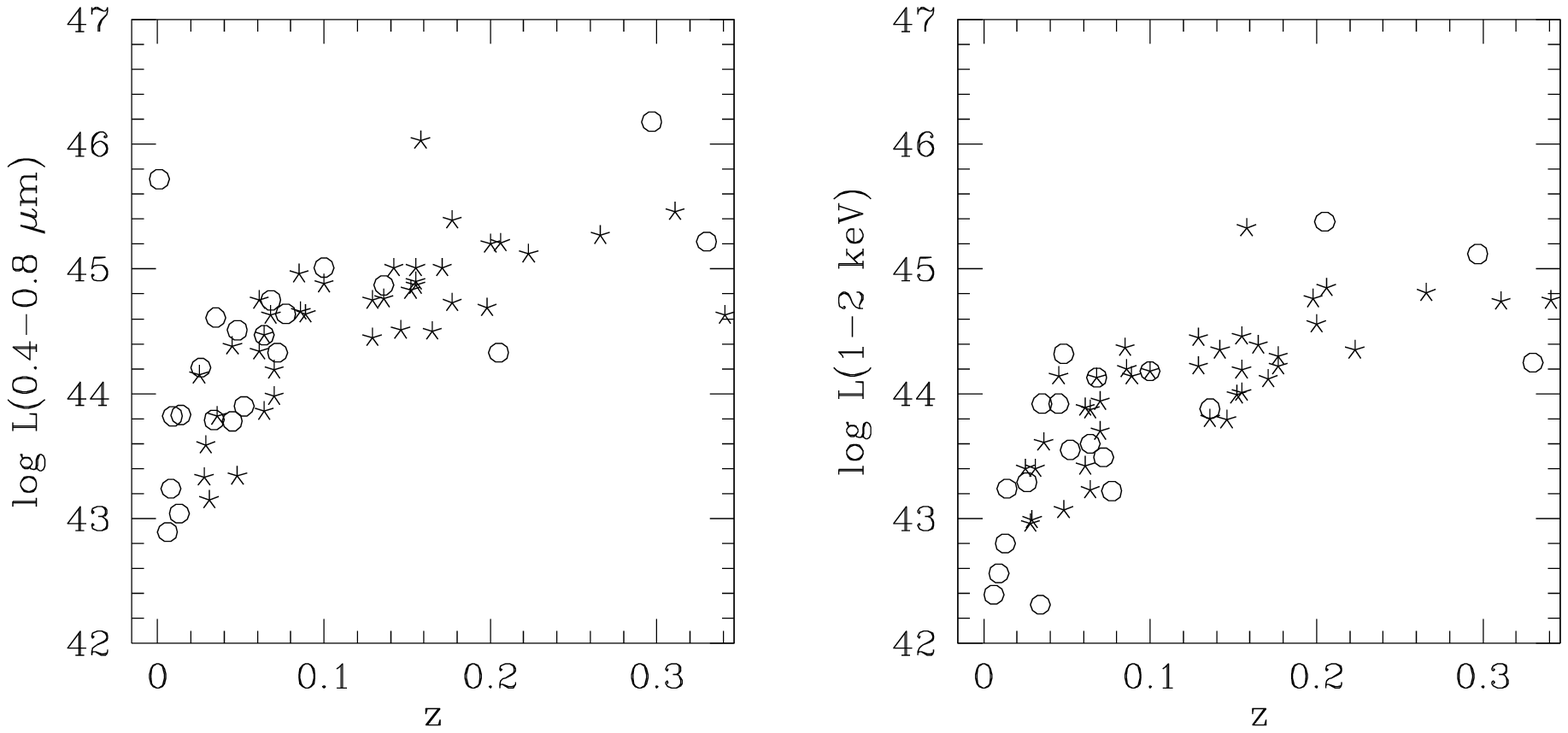}
\caption{Relation between luminosity and redshift. Open circles
indicate our sample, stars Elvis et al. (1994) sample.}
\label{fig:L_z} 
\end{figure}

\begin{figure}
\plotone{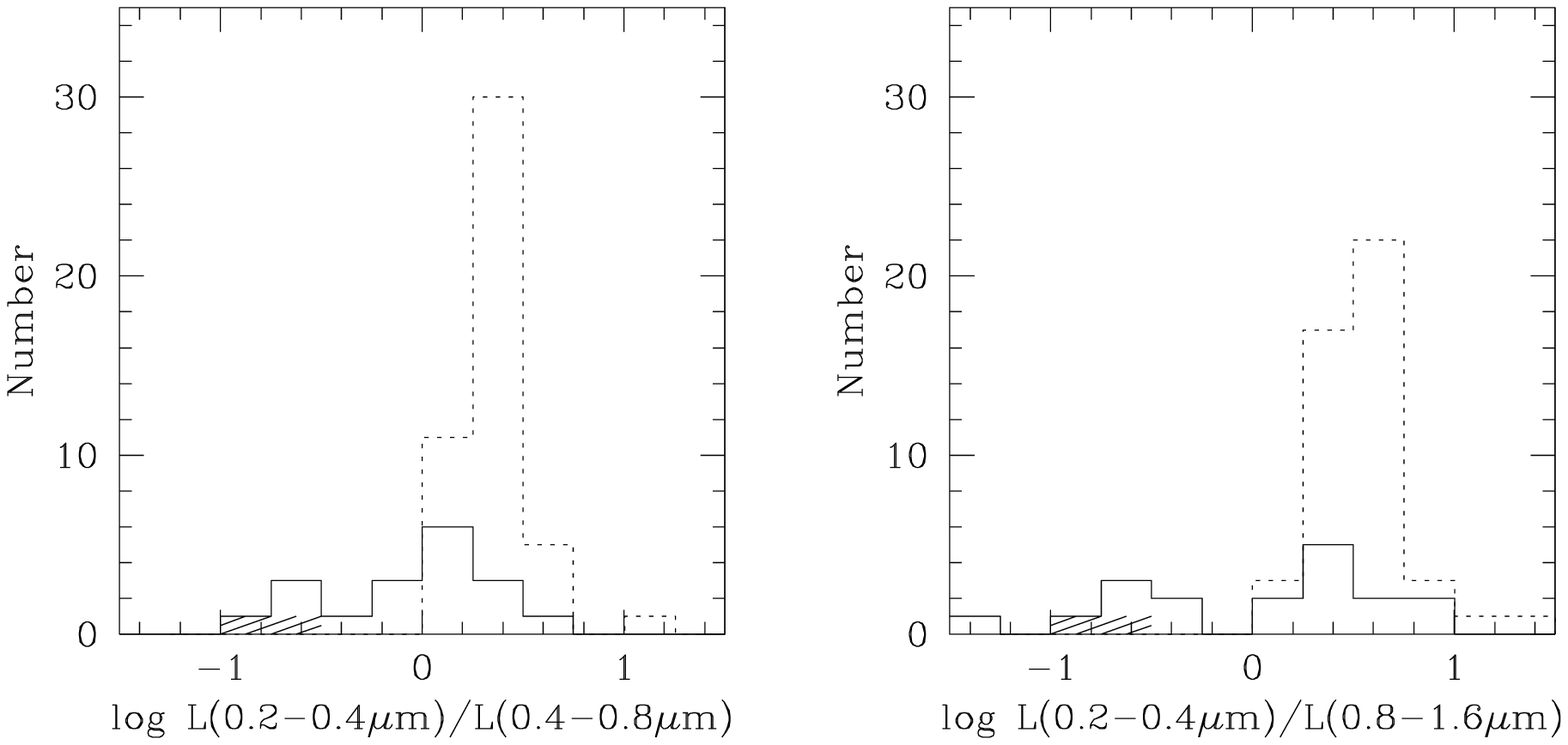}
\caption{Histograms of L(0.2-0.4$\mu$m)/L(0.4-0.8$\mu$m) and
L(0.2-0.4$\mu$m)/L(0.8-1.6$\mu$m). Our sample is represented by a
solid line, while the Elvis et al. (1994) sample - by a dotted line
respectively. We have excluded 3A~05574$-$383, H1419+480 and H1537+339
from the histograms, as these objects did not have enough data points
in the 0.2$-$0.4$\mu$m wavelength range to calculate the
luminosity. Hatched regions represent IC~4329A and H1834$-$653 which
have intrinsic $N_H$.}
\label{fig:histo}
\end{figure}

\begin{figure}
\epsscale{0.55}
\plotone{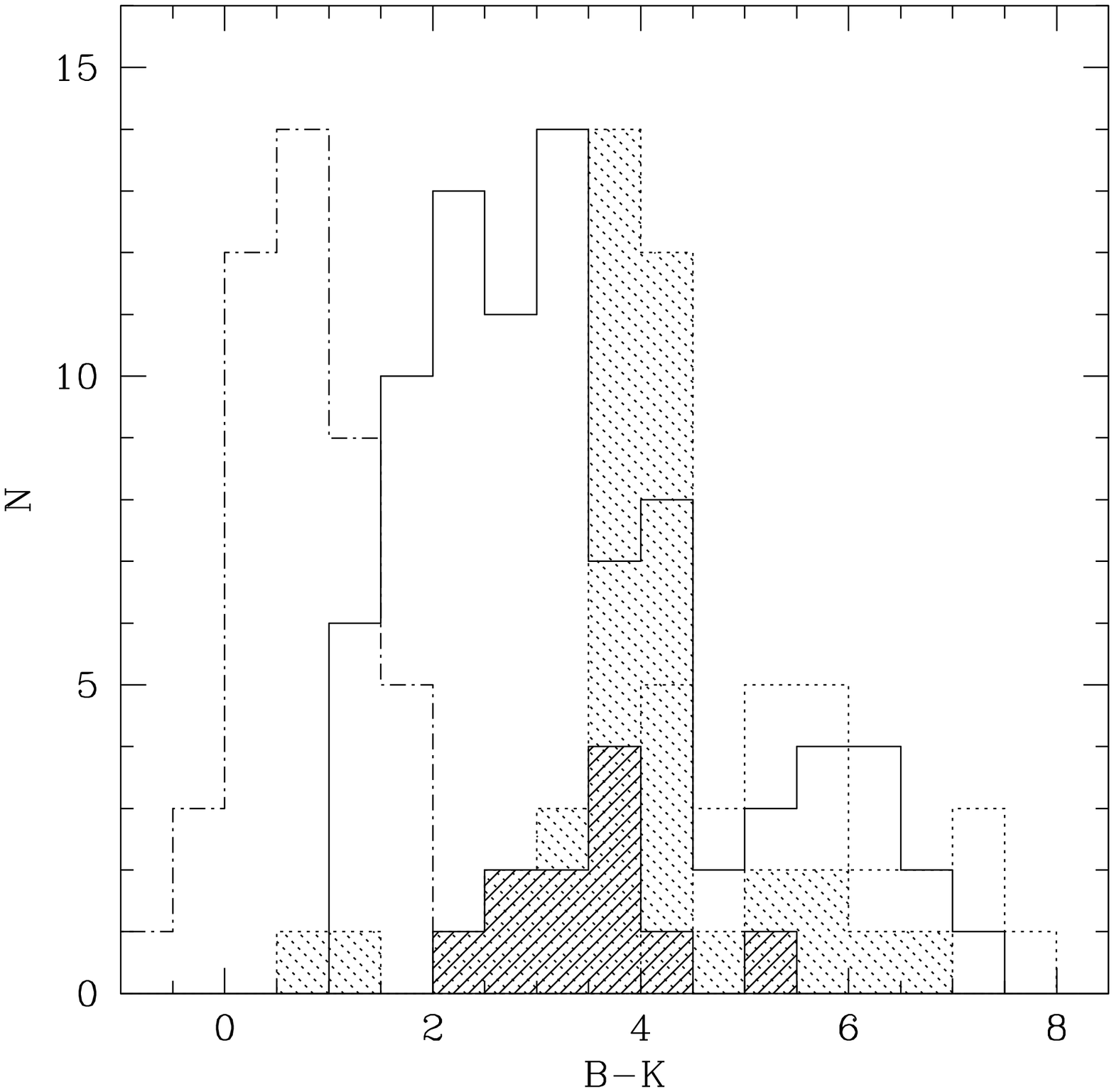}
\caption{Comparison of B-K colors between the HEAO sample
(dotted-line-shaded area), the HEAO sub-sample analyzed in this paper 
(solid-line-shaded area), Webster et al. sample (solid line), PG
sample (dot-dash line), and Chandra observed 2MASS sample (dotted line).}
\label{fig:B_K_mags}
\end{figure}

\begin{figure}
\epsscale{1}
\plotone{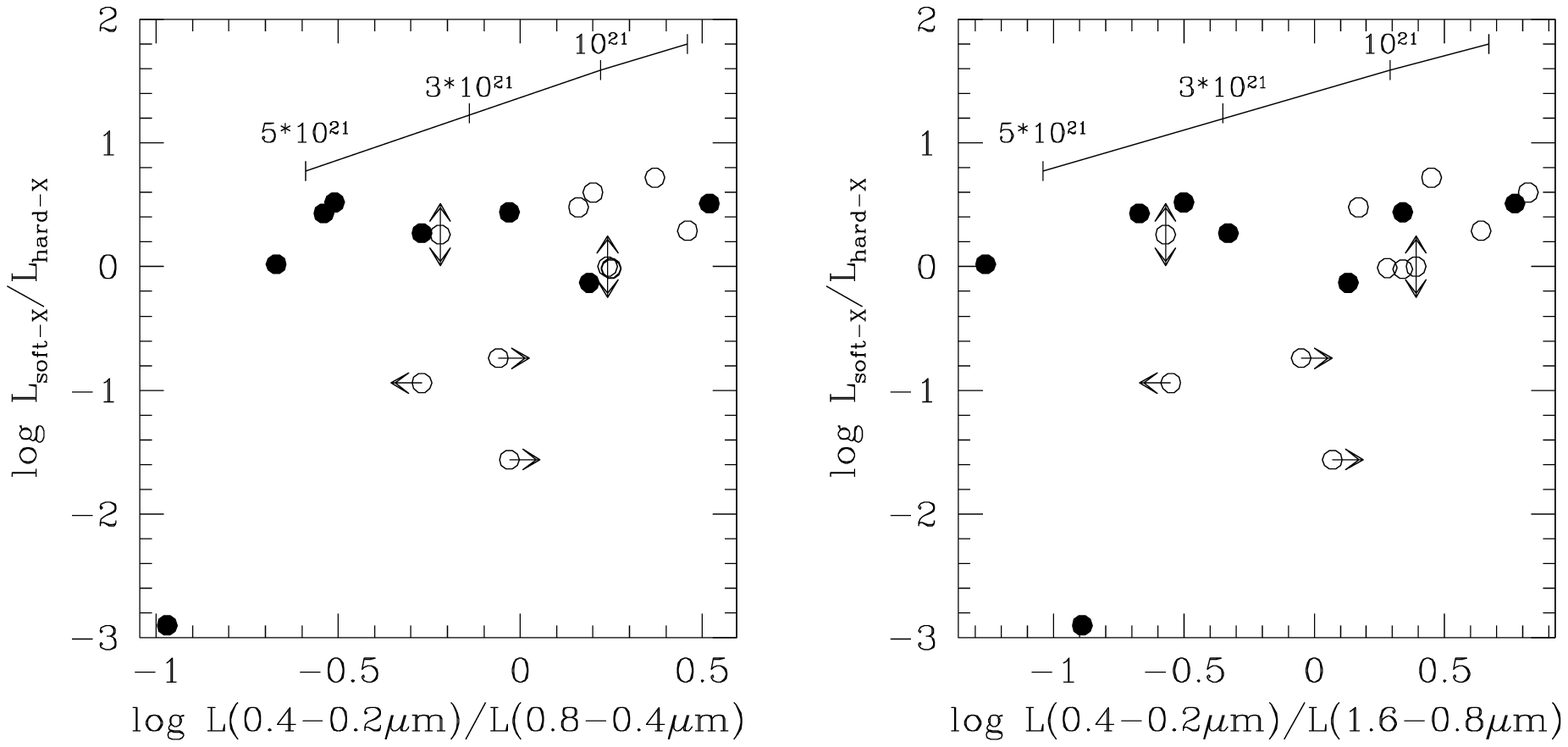}
\caption{X-ray softness ratio vs. L(0.2-0.4$\mu$m)/L(0.4-0.8$\mu$m)
and L(0.2-0.4$\mu$m)/L(0.8-1.6$\mu$m) ratios. Circles with horizontal
arrows represent 3A~05574$-$383, H1419+480 and H1537+339 which do not
have enough data points to calculate the 0.2$-$0.4$\mu$m luminosity so
their position with respect to the X axis is not well
defined. Vertical arrows indicate objects with uncertain ratios due to
lack of soft-X-ray data. Open circles are Seyfert 1s and filled
circles are intermediate and type 2 Seyferts. The Galactic dust
reddening curve for different column densities $N_H$ is shown.} 
\label{fig:04_02_hardness}
\end{figure}

\begin{figure}
\epsscale{0.5}
\plotone{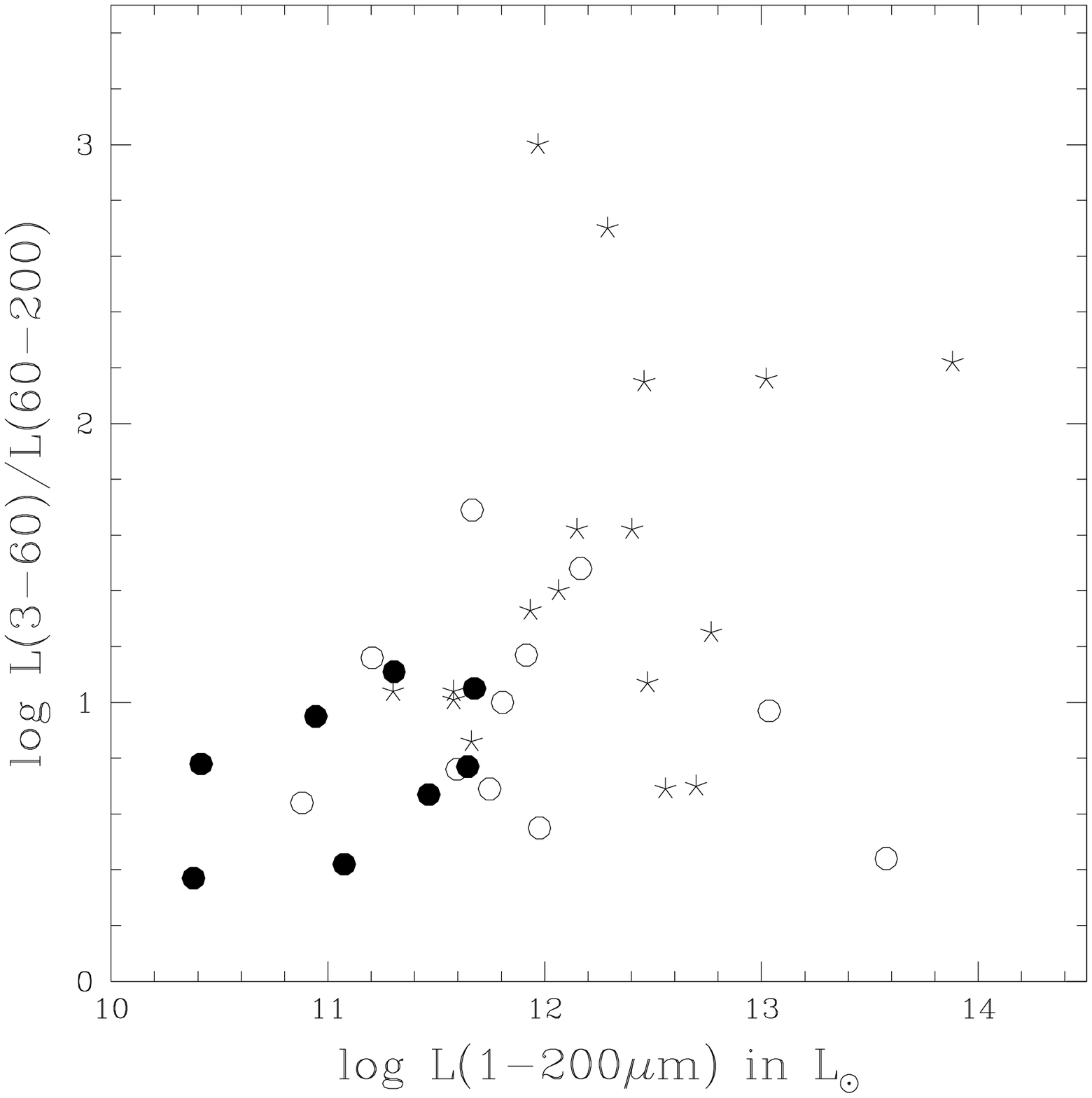}
\caption{The ratio of warm ($3-60 \mu$m) to cool ($60-200\mu$m) IR
luminosity plotted as a function of the total IR luminosity for the
the HEAO sample (filled circles: intermediate and type 2, open circles
type 1) and the Polletta et al. sample of type 1 AGN (stars).}
\label{fig:Poletta}
\end{figure}

\begin{figure}
\plotone{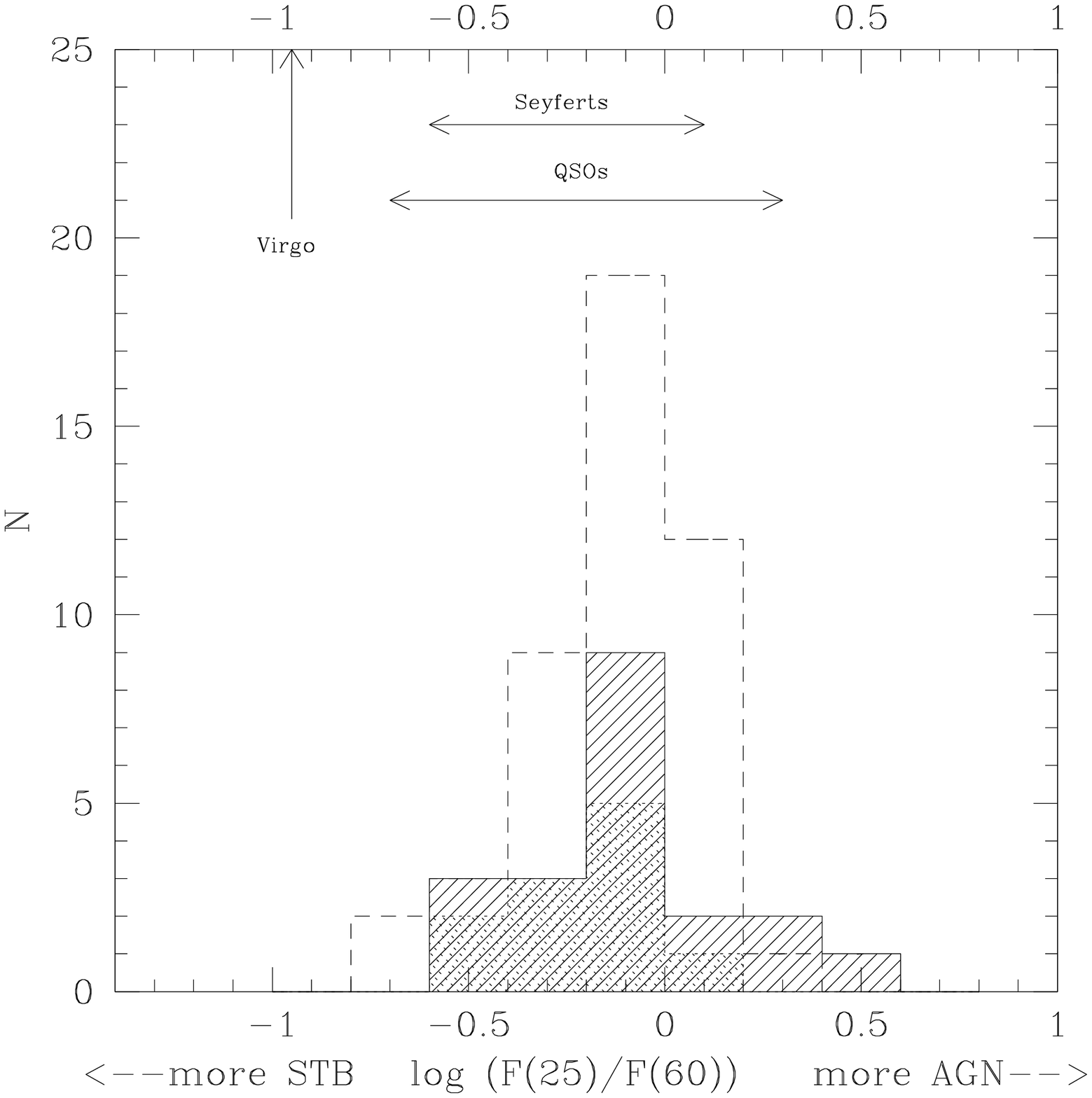}
\caption{Comparison of the distribution of the 25\micron\ to
60\micron\ flux ratio of the HEAO sample (shaded histogram; reddest
objects with L(0.2-0.4$\mu$m)/L(0.4-0.8$\mu$m)$<$0 are indicated by
dotted-line-shaded areas), E94 sample (dashed line histogram), and the
range presented in Seyfert 1 and 2 galaxies from de Grijp et
al. (1985), Low et al. (1988), and the Palomar-Green QSOs from Sanders
et al. (1989). The 25/60 ratio for normal spiral galaxies from the
Virgo cluster from Soifer et al. (1987) is indicated by ``Virgo''.}
\label{fig:alpha_25_60} 
\end{figure}

\begin{figure}
\epsscale{1}
\plotone{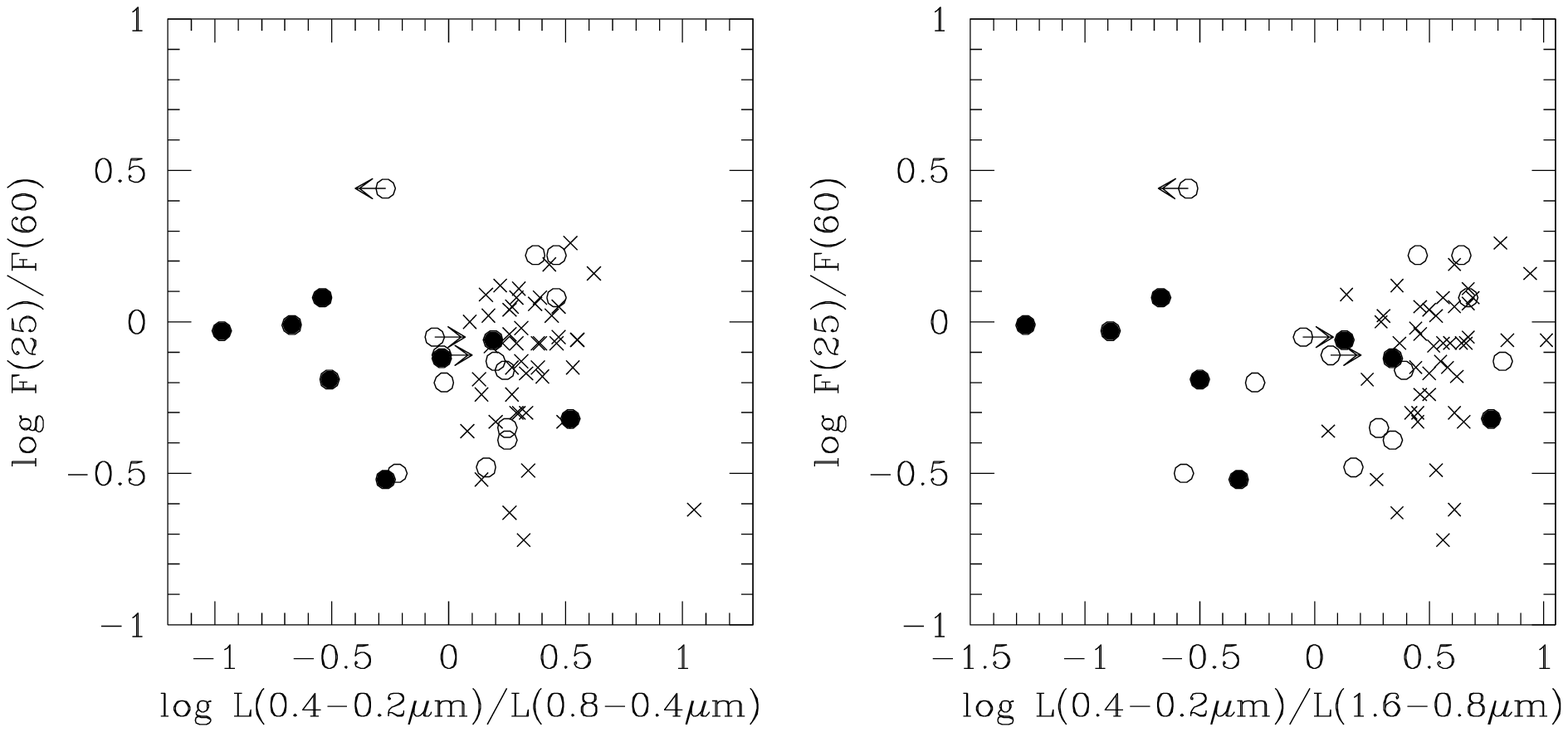}
\caption{The 25\micron\ to 60\micron\ flux ratios versus the
UV/optical reddening indicators. Open circles are HEAO Seyfert 1s,
filled circles HEAO Seyfert 1.2, 1.5, and 2, and crosses E94 quasars.}
\label{fig:25_60tooptUV}
\end{figure}

\begin{figure}
\epsscale{0.5}
\plotone{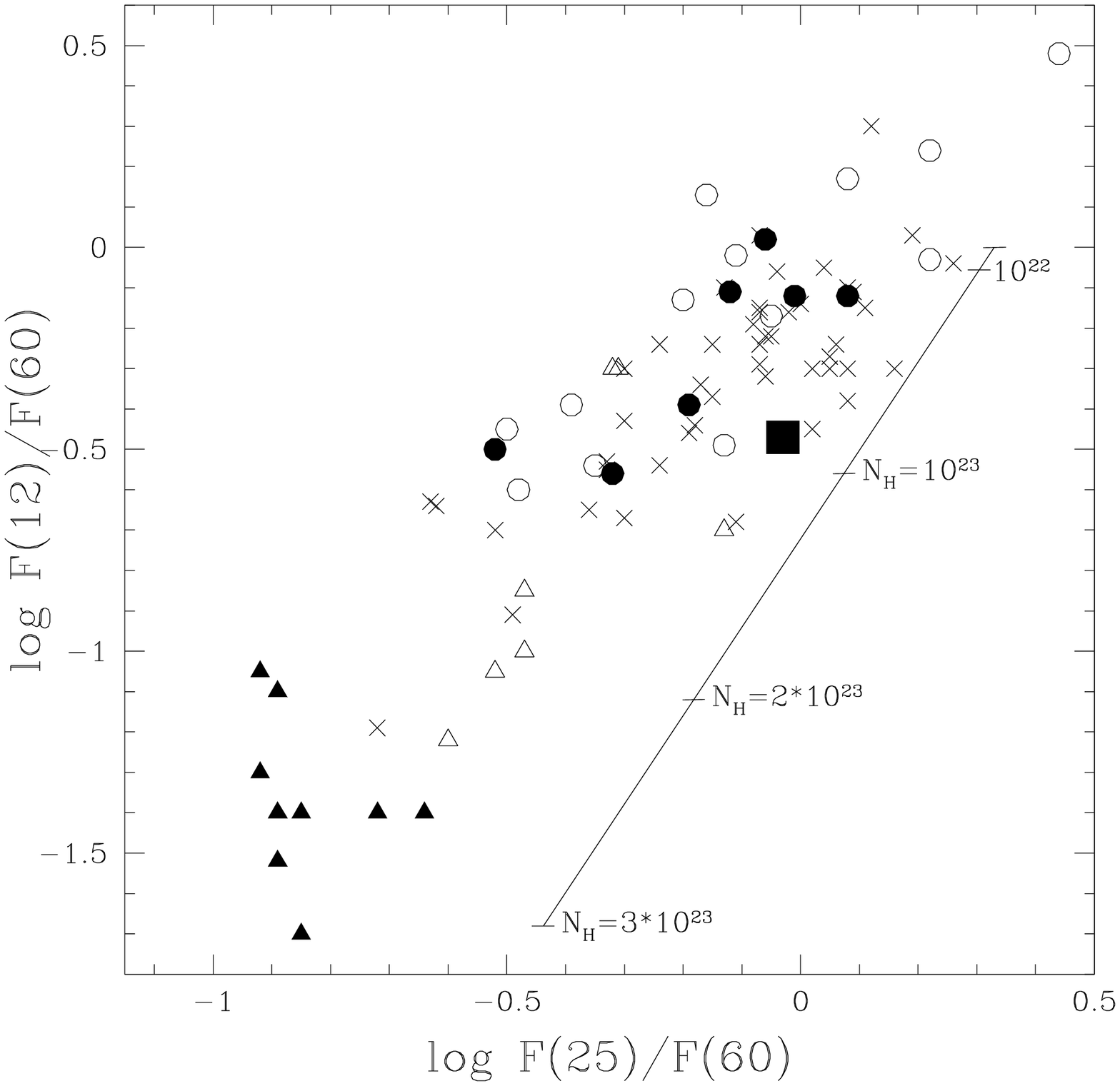}
\caption{The 12/60 versus 25/60 flux ratios. Sy1s from the HEAO
sample are represented by open circles,  intermediate type Seyferts 
by filled circles, and the type 2 Seyfert by a filled square. Filled 
crosses show the E94 sample, open triangles - Sy2 with a hidden broad
line region, and filled triangles - Sy2 without a hidden broad line 
region from Heisler et al. 1997.}
\label{fig:12_60to25_60}
\end{figure}

\begin{figure}
\epsscale{0.85}
\plotone{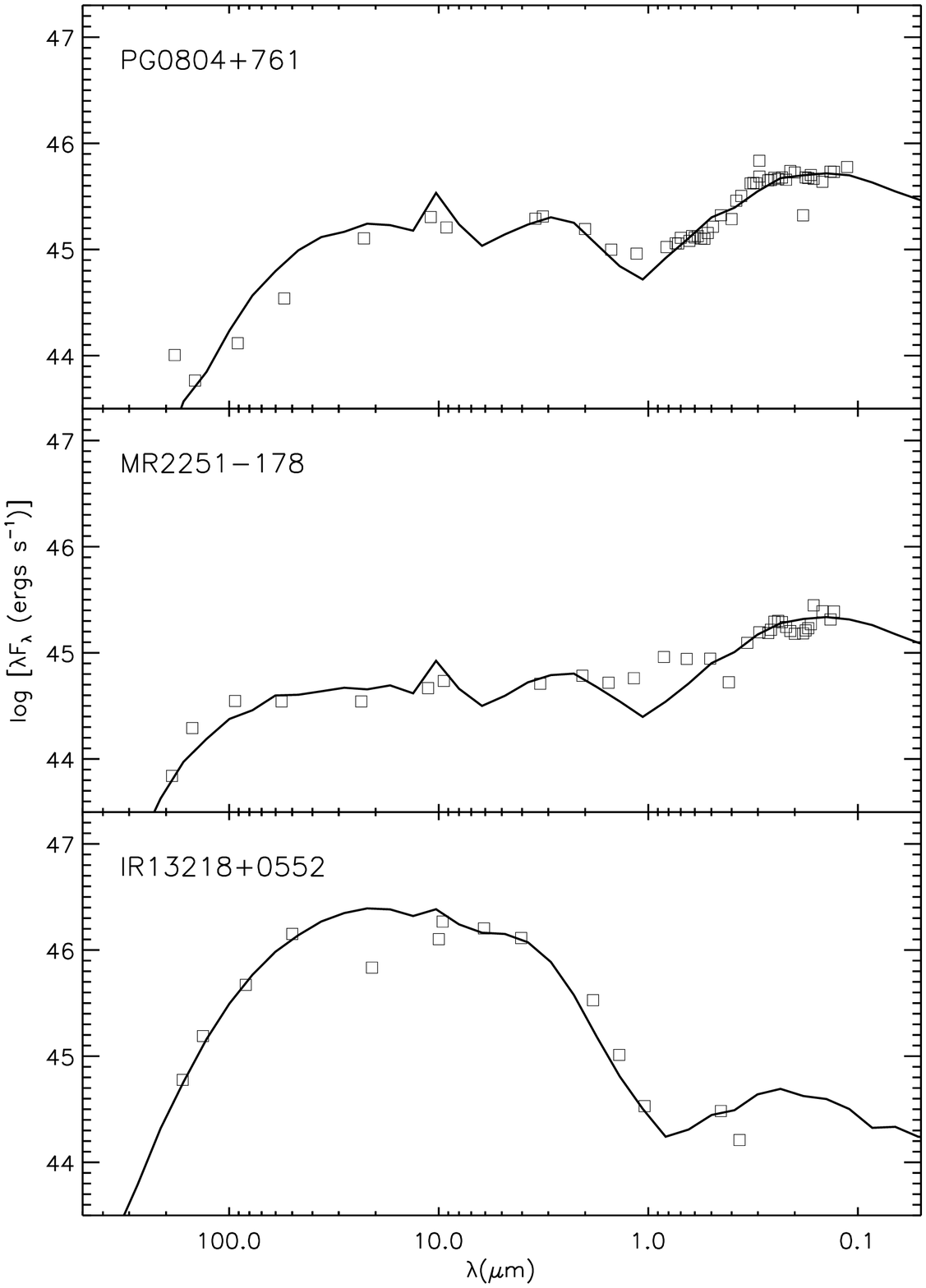}
\vspace{1cm}
\caption{Observed SEDs for PG~0804+761, MR~2251$-$178 and IR~13218+0552
compared with the model SED as described in the text (Section~4.3) and
Table~8.}  
\label{fig:kw1}
\end{figure}

\begin{figure}
\epsscale{0.85}
\plotone{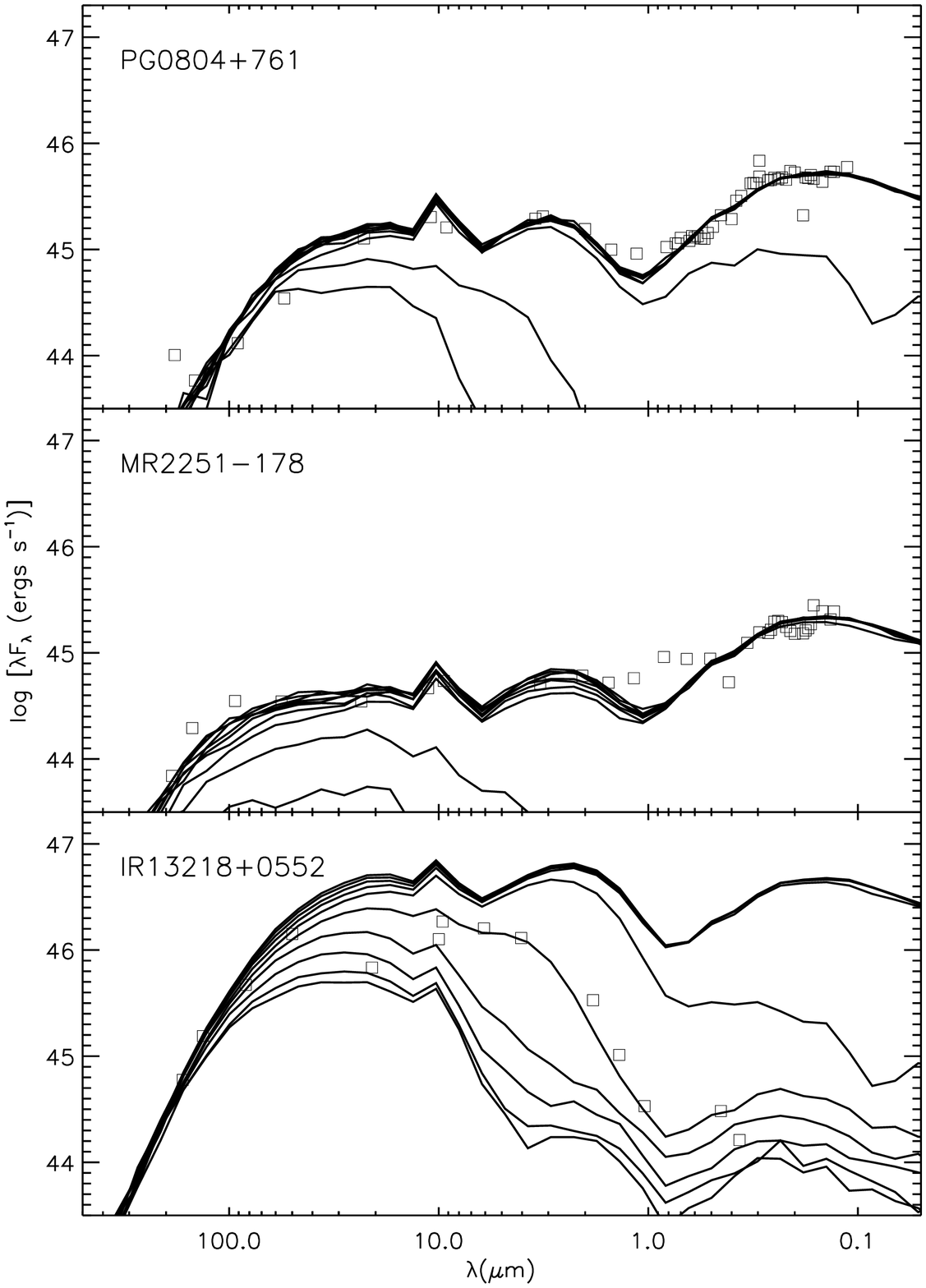}
\vspace{1cm}
\caption{Observed SEDs for PG~0804+761, MR~2251$-$178 and IR~13218+0552
superposed on their best fit model as viewed at a range  of
10 inclinations evenly spaced in $cos(i)$ (starting from $cos(i)=0.05$
to 0.95).}
\label{fig:kw2}
\end{figure}

\clearpage

\begin{figure}
\plotone{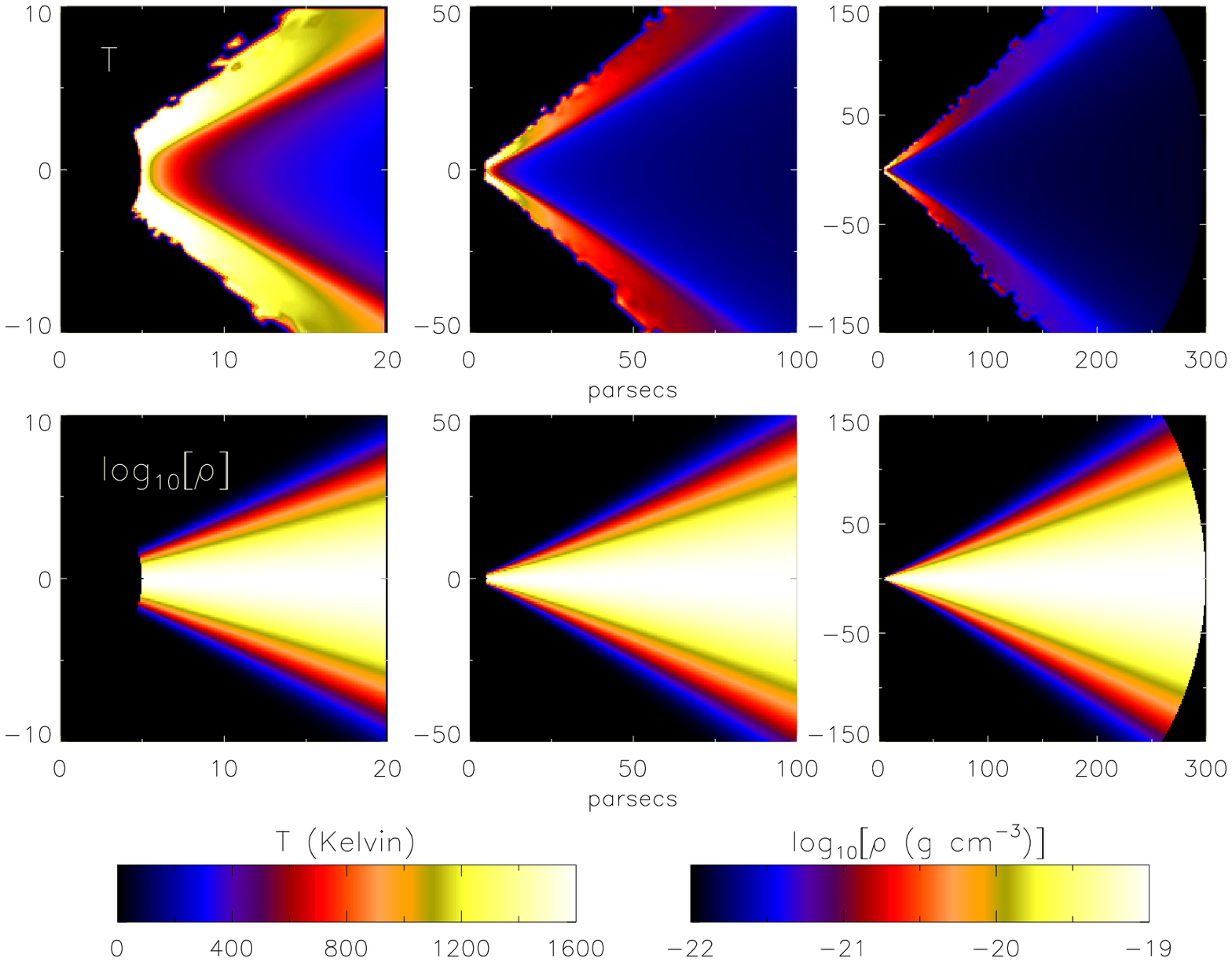}
\vspace{1cm}
\caption{Temperature (top row) and density (bottom row) distributions 
of dust in the circumnuclear dusty disk-like structure. For details
see section~4.3.} 
\label{fig:kw3}
\end{figure}


\clearpage
\begin{deluxetable}{lrlll}
\tablenum{1}
\tablecaption{Sample}
\tablewidth{0pt}
\tablehead{
\colhead{Name} &
\colhead{z} &
\colhead{type} &
\colhead{$\alpha$(J2000)} &
\colhead{$\delta$(J2000)} \\
}
\startdata
MKN 1152&	0.053 & Sy1.5 & 01$^h$ 13$^m$ 50$^s$.09&          $-$14$^o$ 50' 46.''5 \\
MKN 590&	0.026 & Sy1.2 & 02$^h$ 14$^m$ 33$^s$.56&          $-$00$^o$ 45' 59.''9 \\
ESO 198-G24 &	0.046 & Sy1   & 02$^h$ 38$^m$ 19$^s$.52&          $-$52$^o$ 11' 31.''5 \\
3A 0557$-$383  &0.034 & Sy1   & 05$^h$ 58$^m$ 02$^s$.92&          $-$38$^o$ 20' 05.''7 \\
PG~0804+761&	0.100 & Sy1   & 08$^h$ 10$^m$ 56$^s$.31&            +76$^o$ 02' 43.''0  \\
H~1039$-$074&	0.674 & \ldots& 10$^h$ 42$^m$ 19$^s$.16&	      $-$07$^o$ 40' 35.''4\\
NGC 3783&	0.010 & Sy1   & 11$^h$ 39$^m$ 01$^s$.77&          $-$37$^o$ 44' 19.''7 \\
TON 1542&	0.063 & Sy1   & 12$^h$ 32$^m$ 03$^s$.60&            +20$^o$ 09' 30.''0\\
IRAS 13218+0552&0.205 & Sy1   & 13$^h$ 24$^m$ 21$^s$.58&            +05$^o$ 36' 57.''1\\
MCG $-$6$-$30$-$15 & 0.008 & Sy1.2 & 13$^h$ 35$^m$ 53$^s$.93&     $-$34$^o$ 17' 42.''5\\
IC 4329A&	0.016 & Sy1.2 & 13$^h$ 49$^m$ 19$^s$.39&          $-$30$^o$ 18' 35.''3 \\
H~1419+480&	0.072 & Sy1   & 14$^h$ 21$^m$ 29$^s$.42&            +47$^o$ 47' 27.''8 \\
MKN~478 &	0.079 & Sy1   & 14$^h$ 42$^m$ 07$^s$.47&            +35$^o$ 26' 23.''0 \\ 
H~1537+339&	0.330 & Sy1   & 15$^h$ 39$^m$ 52$^s$.23&            +33$^o$ 49' 31.''1 \\
KAZ 102&	0.136 & Sy1   & 18$^h$ 03$^m$ 28$^s$.80 &           +67$^o$ 38' 10.''0  \\
E~1821+643&	0.297 & Sy1   & 18$^h$ 21$^m$ 54$^s$.89&            +64$^o$ 21' 12.''1  \\
H~1834$-$653&	0.013 & Sy2   & 18$^h$ 38$^m$ 20$^s$.28 &         $-$65$^o$ 25' 41.''8 \\
MKN 509&	0.034 & Sy1.2 & 20$^h$ 44$^m$ 09$^s$.74 &         $-$10$^o$ 43' 24.''5  \\
NGC 7213&	0.006 & Sy1.5 & 22$^h$ 09$^m$ 16$^s$.58&          $-$47$^o$ 09' 36.''0 \\
MR 2251$-$178&	0.064 &Sy1    & 22$^h$ 54$^m$ 05$^s$.10 &         $-$17$^o$ 34' 47.''0 \\
MCG $-$2$-$58$-$22 &0.047 & Sy1.5 & 23$^h$ 04$^m$ 43$^s$.48 &     $-$08$^o$ 41' 08.''7\\
\enddata
\end{deluxetable}

\clearpage
\begin{deluxetable}{lccccccccccc}
\rotate
\tabletypesize{\scriptsize}
\tablenum{2}
\tablecaption{ISO Fluxes$^a$}
\tablewidth{0pt}
\tablehead{
\colhead{Name} &
\colhead{5$\mu$m} &
\colhead{7$\mu$m} &
\colhead{12$\mu$m} &
\colhead{25$\mu$m} &
\colhead{C or R$^b$} &
\colhead{60$\mu$m} &
\colhead{100$\mu$m} &
\colhead{C or R$^b$} &
\colhead{135$\mu$m} &
\colhead{200$\mu$m} &
\colhead{C or R} \\
}
\startdata
MKN 1152&\ldots&\ldots&\ldots&\ldots&\ldots&0.213$\pm$0.022&0.445$\pm$0.017&R&0.547$\pm$0.030&0.321$\pm$0.041&R\\
MKN 590&     0.019$\pm$0.003 &0.322$\pm$0.003 &0.288$\pm$0.005 &0.390$\pm$0.013&R&0.432$\pm$0.020&1.135$\pm$0.036&R&1.802$\pm$0.053&1.609$\pm$0.064&R\\
ESO 198$-$G24 &\ldots&0.058$\pm$0.003&0.071$\pm$0.005&0.082$\pm$0.013&R&0.107$\pm$0.020 &0.128$\pm$0.023&R&0.051$\pm$0.031&0.009$\pm$0.040&R\\
3A 0557-383  & 0.011$\pm$0.003 &0.220$\pm$0.003 &0.348$\pm$0.005 &0.505$\pm$0.013 &R&0.221$\pm$0.036&0.151$\pm$0.046 &R&0.172$\pm$0.014&0.117$\pm$0.040&R\\
PG~0804+761&  0.020$\pm$0.003 &0.217$\pm$0.003 &0.195$\pm$0.005&0.246$\pm$0.013&R&0.145$\pm$0.020&0.097$\pm$0.024 &R&0.072$\pm$0.018&0.051$\pm$0.040&R\\
H~1039-074&$<$0.006&0.014$\pm$0.003 &0.086$\pm$0.020 &0.053$\pm$0.019&C&$<$0.060&$<$0.090 &C&$<$0.150&$<$0.210&C\\
NGC~3783& \ldots &0.487$\pm$0.015 &1.976$\pm$0.031 &2.934$\pm$0.097 &C&\ldots& \ldots&\ldots&3.910$\pm$0.071&1.949$\pm$0.182&C\\ 
TON 1542&    0.021$\pm$0.003 &0.045$\pm$0.010 &0.170$\pm$0.096 &0.190$\pm$0.089&C&\ldots&\ldots &\ldots&0.111$\pm$0.036&0.105$\pm$0.051&C\\
IRAS 13218+0552&0.100$\pm$0.015 &0.189$\pm$0.014 &0.338$\pm$0.076 &0.313$\pm$0.015&C&3.01$\pm$1.2&0.53$\pm$0.21&C&0.270$\pm$0.041&$<$0.180&C\\
MCG $-$6$-$30$-$15&0.126$\pm$0.027 &0.238$\pm$0.015 &0.687$\pm$0.013 &0.843$\pm$0.319&C&\ldots&\ldots&\ldots&0.584$\pm$0.060&0.096$\pm$0.100 &C\\
IC 4329A&\ldots &1.221$\pm$0.020 &3.089$\pm$0.321 &2.151$\pm$0.077 &C&\ldots&1.384$\pm$0.025      &R&1.161$\pm$0.040 &0.763$\pm$0.047&R\\
H~1419+480& 0.018$\pm$0.012 &0.099$\pm$0.015 &0.276$\pm$0.029 &0.260$\pm$0.175&C&$<$0.603&$<$0.589 &C& 0.124$\pm$0.051&$<$0.210&C\\
MKN 478 &   0.037$\pm$0.003 &0.061$\pm$0.005 &0.121$\pm$0.033 &0.218$\pm$0.049 &C&\ldots&\ldots  &C&0.457$\pm$0.046&$<$0.600&C\\ 
H1537+339& 0.004$\pm$0.003&0.006$\pm$0.003 & \ldots&0.070$\pm$0.019&C&$<$0.150&$<$0.120    &C&0.178$\pm$0.031&0.170$\pm$0.060&C\\
KAZ 102&0.038$\pm$0.010 &0.025$\pm$0.015&\ldots&\ldots&R& $<$0.075 &0.038$\pm$0.010&R&0.152$\pm$0.061&0.166$\pm$0.085&C\\
E1821+643&\ldots & 0.195$\pm$0.013&0.700$\pm$0.107&0.745$\pm$0.460&C&\ldots&\ldots    &\ldots&\ldots&\ldots&\ldots\\
H~1834$-$653&\ldots &\ldots&\ldots&\ldots&\ldots&\ldots    &\ldots&\ldots& 0.600$\pm$ 0.051&0.306$\pm$0.037&R\\
MKN 509&  0.128$\pm$0.008 &0.334$\pm$0.003 &0.847$\pm$0.102 &0.586$\pm$0.164&C&\ldots&\ldots  &\ldots&\ldots&0.225$\pm$0.147&C\\
NGC 7213&\ldots   &\ldots&\ldots&\ldots&\ldots&\ldots    &3.67$\pm$0.142&R&6.772$\pm$0.340&5.463$\pm$0.392&R\\
MR 2251$-$178& 0.092$\pm$0.005 &0.133$\pm$0.005 &0.300$\pm$0.070 &0.152$\pm$0.015&C&\ldots &\ldots  &\ldots&0.508$\pm$0.042&0.225$\pm$0.061&C\\
MCG $-$2$-$58$-$22 & 0.112$\pm$0.021 &0.176$\pm$0.007 &0.392$\pm$0.220 &0.238$\pm$0.029&C&\ldots &\ldots  &C&0.332$\pm$0.050&0.154$\pm$0.066&C\\
\enddata
\tablenotetext{a}{Fluxes are in Jy}
\tablenotetext{b}{C -- chopped data, R -- raster scans} 
\end{deluxetable}

\clearpage
\begin{deluxetable}{lccccccccccc}
\rotate
\tabletypesize{\scriptsize}
\tablewidth{0pt}
\tablenum{3}
\tablecaption{Spectral energy distributions details for HEAO sample}
\tablehead{
\colhead{Name} &
\colhead{radio} &
\colhead{sub$-$mm} &
\colhead{far$-$IR} &
\colhead{Opt/near-IR} &
\colhead{UV} &
\colhead{X} &
\colhead{Starlight$^{a}$} & 
\colhead{Half-light radius} &
\colhead{Ref.} &
\colhead{Gal. $N_H$$^b$} & 
\colhead{Intrinsic absorption? $^c$}\\
\colhead{1}&
\colhead{2}&
\colhead{3}&
\colhead{4}&
\colhead{5}&
\colhead{6}&
\colhead{7}&
\colhead{8}&
\colhead{9}&
\colhead{10}&
\colhead{11}&
\colhead{12}\\
}
\startdata
MKN~1152   &\ldots  &\ldots     &9,30    &9,15,25,31&    &17,22   &43.87                  &\ldots  & 34  &1.67&no \\     
MKN~590    &\ldots   &\ldots     &25,30   &14,25     &23  &2,22,23 &44.10                  &\ldots  &36,37&2.60&no \\     
ESO 198$-$G24&\ldots   &\ldots     &9,30    &25        &24  &24      &43.26                  & \ldots & 34  &5.41&no \\     
3A 0557$-$383&\ldots   &\ldots     &9,30    &25	       &\ldots     &17,20,22&44.60$^{+0.30}_{-0.60}$&10.0& 34  &3.38&70$^{+90}_{-40}$ \\     
PG~0804+761&13&\ldots     &9,21,30 &16,7      &7  &7,27    &44.38$^{+0.12}_{-0.12}$& 7.5& 36  &3.09&no \\     
H~1039$-$074&\ldots   & \ldots    &9,30    &9      &\ldots     &28      &44.60$^{+0.30}_{-0.60}$&10.0&...  &3.75&? \\                       
NGC~3783   &\ldots   &\ldots     &9,30    &9,19      &\ldots     &17,22   &43.50                  &\ldots  & 34  &9.01&no \\     
TON 1542    &13&\ldots     &9,21,30 &6,9,16    &7  &2,4     &44.41$^{+0.12}_{-0.12}$&7.5 &32,36&2.58&no \\     
IRAS 13218+0552&\ldots   &\ldots     &9,5,30  &5,9       & \ldots    &18	     &44.60$^{+0.05}_{-0.60}$&10.0&...  &2.30&? \\                       
MCG $-$6$-$30$-$15& \ldots &\ldots    &9,25,30 &9         &23  &17,23   &43.00                  &\ldots & 34  &4.06&no \\                       
IC~4329A    &\ldots  &\ldots    &9,30    &9,19,25   &\ldots    &17,22   &42.62                  &... & 34  &4.55&15.0$^{+4}_{-3}$ \\     
H~1419+480   & \ldots &\ldots    &9,30    &9      &\ldots    &2,18    &44.60$^{+0.10}_{-0.60}$&10.0&...  &1.72&no \\     
MKN 478   &13&3   &9,30    &9,16,19   &8,23&1,2,23  &44.31$^{+0.12}_{-0.12}$&6.25& 36  &1.01&no \\     
H 1537+339   & \ldots & \ldots   &9,30    &18	       &\ldots    &18,29   &44.60$^{+0.30}_{-0.60}$&10.0&...  &1.97&no \\                       
KAZ 102   &11&\ldots    &9,30    &7,9       &7  &2,26    &43.86$^{+0.04}_{-0.04}$&11.9& 32  &4.44&no \\      
E 1821+643   & \ldots &3,10&9,30    &9	       & \ldots   &2,27    &44.60$^{+0.30}_{-0.60}$&10.0& 33  &3.50&no \\     
H 1834$-$653   & \ldots &\ldots    &9,30    &9         &\ldots    &17,22   &43.44                  &\ldots & 34  &6.31&1350$^{+330}_{-230}$\\  
MKN~509    &\ldots  &3,10&9,25,30 &19,15,25  &7  &2,17,22 &44.02                  &... & 35  &3.93&no\\     
NGC~7213   & \ldots &\ldots    &9,30    &9	 &\ldots    &17,22   &43.64                  & \ldots& 34  &3.00&no \\     
MR 2251$-$178   &11&\ldots    &9,30    &12,14,15  &7  &7,22    &44.65$^{+0.04}_{-0.04}$&12.2& 32  &2.82&no \\     
MCG $-$2$-$58$-$22&\ldots  &\ldots    &30      &14,15,25  &23  &17,22,23&44.60$^{+0.30}_{-0.60}$&10.0& 34  &3.47&no \\     
\enddata
\tablenotetext{a}{Log of host galaxy luminosity in H band, where
half-light radius is quoted, otherwise log of luminosity in V
band.}
\tablenotetext{b}{Galactic column in units of $10^{20}$~cm$^{-2}$.}
\tablenotetext{c}{Intrinsic $N_H$ is in units of $10^{20}$~cm$^{-2}$
from X-ray analysis- for references see column~7.}
\tablecomments{{\bf References:}
(1)  Boller, Brandt, \& Fink 1996,
(2)  Ceballos \& Barcons 1996,
(3)  Chini, Kreysa, \& Biermann 1989,
(4)  Comastri et al. 1992,
(5)  Low et al. 1989,
(6)  Cutri et al. 1985,
(7)  Elvis et al. 1994,
(8)  Gondhalekar et al. 1994,
(9)  Grossan 1992
(10) Hughes et al. 1993,
(11) Hutchings \& Gower 1985,
(12) Hyland \& Allen 1982,
(13) Kellerman et al. 1989,
(14) McAlary et al. 1983,
(15) H, J, K photometry - this paper
(16) Neugebauer et al. 1987,
(17) Piccinoti et al. 1982,
(18) Remillard et al. 1993,
(19) Rieke 1978,
(20) Rush  et al. 1996,
(21) Sanders et al. 1989,
(22) Turner \& Pounds 1989,
(23) Walter \& Fink 1993, 
(24) Wang, Lu, \& Zhou  1998,
(25) Ward et al. 1987, 
(26) Wilkes \& Elvis 1987,
(27) Williams et al. 1992,
(28) Wood et al. 1994,
(29) Yuan et al. 1998,
(30) ISO observations - this paper,
(31) MMT optical spectra - this paper, 
(32)  Hutchings, Crampton, \& Campbell 1984,
(33)  Hutchings, Janson, \& Neff 1989,
(34)  Kotilainen, Ward, \& Williger 1993,
(35)  MacKenty 1990,
(36)  McLeod \& Rieke 1994,
(37)  McLeod \& Rieke 1995.}
\end{deluxetable}


\clearpage
\begin{deluxetable}{llll}
\tablenum{4}
\tablecaption{Near-IR Photometry}
\tablewidth{0pt}
\tablehead{
\colhead{Name} &
\colhead{J} & 
\colhead{H} &
\colhead{K} \\
}
\startdata
MKN 1152&       12.63 & 11.90 & 11.37 \\
MKN 509&        11.73 & 10.94 & 10.09 \\
MR 2251$-$178&    12.56 & 11.89 & 11.11 \\
MCG $-$2$-$58$-$22 &  12.27 & 11.56 & 10.94 \\ 
\enddata
\end{deluxetable}

\clearpage
\topmargin 0.7in
\begin{deluxetable}{lrrrrrrrrrrrcc}
\rotate
\tabletypesize{\tiny}
\tablenum{5}
\tablecaption{IR Luminosities$^a$}
\tablewidth{0pt}
\tablehead{
\colhead{Name} &
\colhead{L(1.6-3.2)} &
\colhead{L(3.2-6.4)} &
\colhead{L(6.4-12.8)} &
\colhead{L(12.8-25)} &
\colhead{L(25-50)} &
\colhead{L(50-100)} &
\colhead{L(1-10)} &
\colhead{L(10-100)} &
\colhead{L(3-60)} &
\colhead{L(60-200)} &
\colhead{L(1-200)} &
\colhead{$log(\frac{F(25)}{F(60)})$}&
\colhead{$\alpha$(25/60)}\\
}
\startdata
MKN 1152&
$  44.17 ^{+   0.11 }_{-   0.11 } $ &
$  44.17 ^{+   0.09 }_{-   0.11 } $ &
$  44.33 ^{+   0.10 }_{-   0.13 } $ &
$  44.33 ^{+   0.08 }_{-   0.10 } $ &
$  44.13 ^{+   0.12 }_{-   0.15 } $ &
$  44.00 ^{+   0.17 }_{-   0.19 } $ &
$  44.73 ^{+   0.11 }_{-   0.12 } $ &
$  44.72 ^{+   0.11 }_{-   0.13 } $ &
$  44.87 ^{+   0.10 }_{-   0.12 } $ &
$  44.20 ^{+   0.12 }_{-   0.12 } $ &
$  45.06 ^{+   0.11 }_{-   0.12 } $ &
-0.01 &
-0.03 \\
MKN 590&
$  43.69 ^{+   0.02 }_{-   0.03 } $ &
$  43.88 ^{+   0.11 }_{-   1.06 } $ &
$ <  44.12 $ &
$ <  44.02 $ &
$ <  43.86 $ &
$  43.75 ^{+   0.09 }_{-   0.20 } $ &
$  44.29 ^{+   0.12 }_{-   0.41 } $ &
$  44.27 ^{+   0.20 }_{-   0.72 } $ &
$  44.43 ^{+   0.20 }_{-   1.28 } $ &
$  44.02 ^{+   0.06 }_{-   0.07 } $ &
$  44.67 ^{+   0.14 }_{-   0.40 } $ &
-0.12&
-0.32 \\
ESO 198-G24&
$  44.16 ^{+   0.02 }_{-   0.02 } $ &
$  44.13 ^{+   0.06 }_{-   0.08 } $ &
$  44.06 ^{+   0.13 }_{-   0.20 } $ &
$  43.87 ^{+   0.11 }_{-   0.18 } $ &
$ <  43.70 $ &
$ <  43.75 $ &
$  44.63 ^{+   0.05 }_{-   0.29 } $ &
$  44.29 ^{+   0.11 }_{-   0.44 } $ &
$  44.59 ^{+   0.09 }_{-   0.12 } $ &
$  43.43 ^{+   0.42 }_{-   0.19 } $ &
$ 44.80 ^{+   0.09 }_{-   0.15 } $ &
-0.20&
-0.53 \\
3A 0557-383  &
$  44.48 ^{+   0.09 }_{-   0.10 } $ &
$  44.65 ^{+   0.12 }_{-   0.13 } $ &
$  44.64 ^{+   0.07 }_{-   0.08 } $ &
$  44.54 ^{+   0.06 }_{-   0.06 } $ &
$  44.17 ^{+   0.07 }_{-   0.07 } $ &
$  43.51 ^{+   0.11 }_{-   0.13 } $ &
$  45.04 ^{+   0.11 }_{-   0.12 } $ &
$  44.85 ^{+   0.06 }_{-   0.07 } $ &
$  45.16 ^{+   0.09 }_{-   0.09 } $ &
$  43.47 ^{+   0.15 }_{-   0.13 } $ &
$ 45.26 ^{+   0.09 }_{-   0.10 } $ &
0.44&
1.16 \\
PG~0804+761&
$  45.06 ^{+   0.06 }_{-   0.06 } $ &
$  45.11 ^{+   0.05 }_{-   0.05 } $ &
$  45.09 ^{+   0.07 }_{-   0.07 } $ &
$  45.00 ^{+   0.08 }_{-   0.09 } $ &
$  44.68 ^{+   0.10 }_{-   0.12 } $ &
$  44.20 ^{+   0.11 }_{-   0.17 } $ &
$  45.56 ^{+   0.06 }_{-   0.06 } $ &
$  45.32 ^{+   0.08 }_{-   0.10 } $ &
$  45.62 ^{+   0.07 }_{-   0.08 } $ &
$  44.14 ^{+   0.11 }_{-   0.36 } $ &
$ 45.76 ^{+   0.07 }_{-   0.07 } $&
0.22&
0.58\\
H1039-074&
$  45.73 ^{+   0.01 }_{-   0.01 } $ &
$  46.15 ^{+   0.07 }_{-   0.08 } $ &
$  46.59 ^{+   0.06 }_{-   1.14 } $ &
$ <  46.02 $ &
$ <  45.75 $ &
$ <  45.68 $ &
$  46.67 ^{+   0.03 }_{-   0.33 } $ &
$ <  46.46 $ &
$  47.15 ^{+   0.32 }_{-   0.97 } $ &
 ...&
$ 46.41 ^{+   0.50 }_{-   0.06 } $&
-0.16&
-0.42 \\
NGC 3783&
$  43.29 ^{+   0.04 }_{-   0.04 } $ &
$  43.55 ^{+   0.03 }_{-   0.03 } $ &
$  43.66 ^{+   0.04 }_{-   0.04 } $ &
$  43.83 ^{+   0.05 }_{-   0.05 } $ &
$  43.80 ^{+   0.03 }_{-   0.03 } $ &
$  43.61 ^{+   0.02 }_{-   0.02 } $ &
$  43.95 ^{+   0.03 }_{-   0.03 } $ &
$  44.28 ^{+   0.04 }_{-   0.04 } $ &
$  44.35 ^{+   0.04 }_{-   0.04 } $ &
$  43.71 ^{+   0.02 }_{-   0.02 } $ &
$ 44.48 ^{+   0.03 }_{-   0.03 } $&
-0.13&
-0.34\\
TON 1542&
$  44.39 ^{+   0.07 }_{-   0.08 } $ &
$  44.30 ^{+   0.06 }_{-   0.07 } $ &
$  44.40 ^{+   0.10 }_{-   0.13 } $ &
$  44.58 ^{+   0.15 }_{-   0.17 } $ &
$  44.35 ^{+   0.14 }_{-   0.15 } $ &
$  43.85 ^{+   0.32 }_{-   0.13 } $ &
$  44.87 ^{+   0.08 }_{-   0.08 } $ &
$  44.90 ^{+   0.16 }_{-   0.16 } $ &
$  45.04 ^{+   0.12 }_{-   0.14 } $ &
$ <  44.28 $ &
$ 45.19 ^{+   0.13 }_{-   0.14 }$ &
0.22&
0.58\\
IRAS 13218+0552&
$  45.54 ^{+   0.07 }_{-   0.09 } $ &
$  45.96 ^{+   0.15 }_{-   0.23 } $ &
$  46.02 ^{+   0.13 }_{-   0.19 } $ &
$  45.74 ^{+   0.15 }_{-   0.20 } $ &
$  45.87 ^{+   0.20 }_{-   0.26 } $ &
$  45.70 ^{+   0.21 }_{-   0.27 } $ &
$  46.32 ^{+   0.13 }_{-   0.19 } $ &
$  46.32 ^{+   0.18 }_{-   0.22 } $ &
$  46.55 ^{+   0.16 }_{-   0.22 } $ &
$  45.58 ^{+   0.22 }_{-   0.27 } $ &
$ 46.63 ^{+   0.16 }_{-   0.21 } $&
-0.59&
-1.55\\
MCG  -6-30-15&
$  43.19 ^{+   0.05 }_{-   0.06 } $ &
$  43.16 ^{+   0.07 }_{-   0.09 } $ &
$  43.23 ^{+   0.04 }_{-   0.04 } $ &
$  43.23 ^{+   0.06 }_{-   0.06 } $ &
$  43.18 ^{+   0.06 }_{-   0.06 } $ &
$  43.04 ^{+   0.05 }_{-   0.05 } $ &
$  43.72 ^{+   0.05 }_{-   0.53 } $ &
$  43.69 ^{+   0.06 }_{-   0.06 } $ &
$  43.83 ^{+   0.06 }_{-   0.06 } $ &
$  43.05 ^{+   0.06 }_{-   0.05 } $ &
$ 44.02 ^{+   0.05 }_{-   0.23 } $&
-0.19&
-0.50\\
IC 4329A&
$  44.05 ^{+   0.05 }_{-   0.05 } $ &
$  44.11 ^{+   0.06 }_{-   0.07 } $ &
$  44.15 ^{+   0.04 }_{-   0.04 } $ &
$  44.25 ^{+   0.04 }_{-   0.04 } $ &
$  44.02 ^{+   0.03 }_{-   0.03 } $ &
$  43.67 ^{+   0.05 }_{-   0.05 } $ &
$  44.59 ^{+   0.05 }_{-   0.35 } $ &
$  44.58 ^{+   0.03 }_{-   0.04 } $ &
$  44.76 ^{+   0.04 }_{-   0.04 } $ &
$  43.65 ^{+   0.06 }_{-   0.06 } $ &
$ 44.90 ^{+   0.04 }_{-   0.16 } $&
0.08&
0.21\\
H1419+480&
$  44.53 ^{+   0.05 }_{-   0.02 } $ &
$  44.58 ^{+   0.02 }_{-   0.02 } $ &
$  44.79 ^{+   0.09 }_{-   0.11 } $ &
$  44.75 ^{+   0.10 }_{-   0.13 } $ &
$  44.59 ^{+   0.08 }_{-   0.10 } $ &
$  44.40 ^{+   0.07 }_{-   0.83 } $ &
$  45.11 ^{+   0.06 }_{-   0.05 } $ &
$  45.15 ^{+   0.07 }_{-   0.18 } $ &
$  45.31 ^{+   0.08 }_{-   0.10 } $ &
$ <  44.31 $ &
$ 45.40 ^{+   0.11 }_{-   0.08 } $&
-0.05&
-0.13\\
Mkn 478 &
$  44.79 ^{+   0.04 }_{-   0.04 } $ &
$  44.68 ^{+   0.03 }_{-   0.03 } $ &
$  44.69 ^{+   0.08 }_{-   0.09 } $ &
$  44.68 ^{+   0.09 }_{-   0.11 } $ &
$  44.72 ^{+   0.06 }_{-   0.07 } $ &
$  44.76 ^{+   0.05 }_{-   0.06 } $ &
$  45.23 ^{+   0.04 }_{-   0.05 } $ &
$  45.25 ^{+   0.07 }_{-   0.08 } $ &
$  45.34 ^{+   0.06 }_{-   0.07 } $ &
$  44.79 ^{+   0.06 }_{-   0.10 } $ &
$ 45.57 ^{+   0.06 }_{-   0.07 } $&
-0.48&
-1.26\\
H1537+339&
$  45.01 ^{+   0.11 }_{-   0.16 } $ &
$  45.08 ^{+   0.06 }_{-   0.08 } $ &
$  45.68 ^{+   0.10 }_{-   0.52 } $ &
$ <  45.55 $ &
$  45.34 ^{+   0.08 }_{-   1.05 } $ &
$  45.16 ^{+   0.04 }_{-   0.05 } $ &
$  45.78 ^{+   0.10 }_{-   0.41 } $ &
$  45.94 ^{+   0.06 }_{-   0.77 } $ &
$  46.08 ^{+   0.07 }_{-   0.60 } $ &
 ... &
$ 46.18 ^{+   0.08 }_{-   0.57 } $&
-0.11&
-0.29\\
Kaz 102&
$  44.81 ^{+   0.08 }_{-   0.10 } $ &
$  44.83 ^{+   0.04 }_{-   0.05 } $ &
$  44.86 ^{+   0.08 }_{-   0.10 } $ &
$  44.68 ^{+   0.09 }_{-   0.18 } $ &
$ <  44.44 $ &
$  43.97 ^{+   0.07 }_{-   0.19 } $ &
$  45.33 ^{+   0.06 }_{-   0.40 } $ &
$  45.03 ^{+   0.08 }_{-   0.32 } $ &
$  45.35 ^{+   0.07 }_{-   0.16 } $ &
$ <  44.18 $ &
$ 45.51 ^{+   0.07 }_{-   0.15 } $&
0.08&
0.21\\
E1821+643&
$  46.25 ^{+   0.02 }_{-   0.02 } $ &
$  46.35 ^{+   0.03 }_{-   0.03 } $ &
$  46.30 ^{+   0.05 }_{-   0.05 } $ &
$  46.30 ^{+   0.13 }_{-   0.16 } $ &
$  46.30 ^{+   0.16 }_{-   0.19 } $ &
$  46.28 ^{+   0.03 }_{-   0.30 } $ &
$  46.80 ^{+   0.02 }_{-   0.60 } $ &
$  46.82 ^{+   0.11 }_{-   0.19 } $ &
$  46.95 ^{+   0.10 }_{-   0.10 } $ &
$  46.51 ^{+   0.23 }_{-   0.74 } $ &
$ 47.17 ^{+   0.03 }_{-   0.41 } $&
-0.35&
-0.92\\
H1834-653&
$  42.79 ^{+   0.02 }_{-   0.02 } $ &
$  43.21 ^{+   0.03 }_{-   0.03 } $ &
$  43.71 ^{+   0.04 }_{-   0.05 } $ &
$  44.04 ^{+   0.03 }_{-   0.03 } $ &
$  44.03 ^{+   0.02 }_{-   0.02 } $ &
$  43.63 ^{+   0.05 }_{-   0.05 } $ &
$  43.73 ^{+   0.03 }_{-   0.81 } $ &
$  44.45 ^{+   0.03 }_{-   0.03 } $ &
$  44.48 ^{+   0.03 }_{-   0.12 } $ &
$  43.53 ^{+   0.07 }_{-   0.07 } $ &
$ 44.54 ^{+   0.03 }_{-   0.12 } $&
-0.03&
-0.08\\
MKN 509&
$  44.43 ^{+   0.04 }_{-   0.04 } $ &
$  44.46 ^{+   0.03 }_{-   0.03 } $ &
$  44.40 ^{+   0.05 }_{-   0.05 } $ &
$  44.46 ^{+   0.07 }_{-   0.07 } $ &
$  44.44 ^{+   0.09 }_{-   0.11 } $ &
$  44.34 ^{+   0.06 }_{-   0.07 } $ &
$  44.94 ^{+   0.04 }_{-   0.04 } $ &
$  44.94 ^{+   0.07 }_{-   0.08 } $ &
$  45.08 ^{+   0.06 }_{-   0.07 } $ &
$  44.31 ^{+   0.06 }_{-   0.22 } $ &
$  45.24 ^{+   0.06 }_{-   0.08 } $ &
-0.32&
-0.84\\
NGC 7213&
$  43.04 ^{+   0.02 }_{-   0.02 } $ &
$  43.12 ^{+   0.03 }_{-   0.04 } $ &
$  43.19 ^{+   0.04 }_{-   0.05 } $ &
$  43.07 ^{+   0.04 }_{-   0.05 } $ &
$  43.05 ^{+   0.04 }_{-   0.04 } $ &
$  43.16 ^{+   0.09 }_{-   0.09 } $ &
$ <  43.63 $ &
$  43.63 ^{+   0.06 }_{-   0.06 } $ &
$  43.75 ^{+   0.04 }_{-   0.34 } $ &
$  43.38 ^{+   0.10 }_{-   0.09 } $ &
$  43.98 ^{+   0.05 }_{-   0.33 } $&
-0.52&
-1.37\\
MR2251-178&
$  44.59 ^{+   0.11 }_{-   0.10 } $ &
$  44.56 ^{+   0.06 }_{-   0.06 } $ &
$  44.55 ^{+   0.06 }_{-   0.07 } $ &
$  44.41 ^{+   0.10 }_{-   0.13 } $ &
$  44.38 ^{+   0.11 }_{-   0.13 } $ &
$  44.32 ^{+   0.05 }_{-   0.06 } $ &
$  45.10 ^{+   0.09 }_{-   0.09 } $ &
$  44.92 ^{+   0.08 }_{-   0.10 } $ &
$  45.12 ^{+   0.08 }_{-   0.09 } $ &
$  44.43 ^{+   0.04 }_{-   0.04 } $ &
$ 45.34 ^{+   0.09 }_{-   0.09 } $&
-0.36&
-0.95\\
MCG -2-58-22 &
$  44.50 ^{+   0.08 }_{-   0.08 } $ &
$  44.59 ^{+   0.06 }_{-   0.07 } $ &
$  44.62 ^{+   0.05 }_{-   0.05 } $ &
$  44.38 ^{+   0.06 }_{-   0.06 } $ &
$  44.16 ^{+   0.05 }_{-   0.06 } $ &
$  43.91 ^{+   0.06 }_{-   0.06 } $ &
$  45.09 ^{+   0.07 }_{-   0.07 } $ &
$  44.77 ^{+   0.06 }_{-   0.29 } $ &
$  45.09 ^{+   0.05 }_{-   0.13 } $ &
$ <  44.04 $ &
$ 45.27 ^{+   0.07 }_{-   0.14 } $&
-0.06&
-0.16\\
\enddata
\tablenotetext{a}{L($\lambda_1$--$\lambda_2$) indicates logarithm of
luminosity integrated between $\lambda_1$[$\mu$m] and $\lambda_2$[$\mu$m] in
units of erg~s$^{-1}$. Cosmological model with $q_{o} = 0$ and $H_{o}$ = 50~km~s$^{-1}$Mpc$^{-1}$ was used.}
\end{deluxetable}


\clearpage
\begin{deluxetable}{lrrrrrrr}
\rotate
\tabletypesize{\scriptsize}
\tablenum{6}
\tablecaption{Optical, UV and X-ray Luminosities$^a$}
\tablewidth{0pt}
\tablehead{
\colhead{Name} &
\colhead{L(0.2-0.1$\mu$m)} &
\colhead{L(0.4-0.2$\mu$m)} &
\colhead{L(0.8-0.4$\mu$m)} &
\colhead{L(1.6-0.8$\mu$m)} &
\colhead{L(0.1-2.0 keV)}&
\colhead{L(2-10 keV)}&
\colhead{$L_{BOL}$(1mm-10 keV)}\\
}
\startdata
MKN 1152&
$  43.00 ^{+   0.01 }_{-   0.01 } $&
$  42.96 ^{+   0.00 }_{-   0.00 } $&
$  43.63 ^{+   0.00 }_{-   0.00 } $&
$  44.22 ^{+   0.11 }_{-   0.11 } $&
$  44.13 ^{+   0.14 }_{-   0.13 } $&
$  44.11 ^{+   0.11 }_{-   0.14 } $&
$  45.19 ^{+   0.10 }_{-   0.11 } $\\
MKN 590&
$  44.36 ^{+   0.05 }_{-   0.05 } $&
$  44.20 ^{+   0.04 }_{-   0.04 } $&
$  44.23 ^{+   0.01 }_{-   0.02 } $&
$  43.86 ^{+   0.02 }_{-   0.02 } $&
$  44.05 ^{+   0.20 }_{-   0.20 } $&
$  43.61 ^{+   0.14 }_{-   0.14 } $&
$  45.16 ^{+   0.09 }_{-   0.14 } $\\
ESO 198-G24 &
$  43.99 ^{+   0.10 }_{-   0.14 } $&
$  43.76 ^{+   0.04 }_{-   0.06 } $&
$  43.78 ^{+   0.00 }_{-   0.00 } $&
$  44.02 ^{+   0.01 }_{-   0.01 } $&
$  44.77 ^{+   0.04 }_{-   0.03 } $&
 ...&
$  45.29 ^{+   0.07 }_{-   0.11 } $\\
3A 0557-383  &
$  43.26 ^{+   0.08 }_{-   0.13 } $&
$  43.52 ^{+   0.11 }_{-   0.16 } $&
$  43.79 ^{+   0.13 }_{-   0.20 } $&
$  44.07 ^{+   0.14 }_{-   0.21 } $&
$  42.95 ^{+   0.01 }_{-   0.01 } $&
$  43.89 ^{+   0.09 }_{-   0.09 } $&
$  45.31 ^{+   0.09 }_{-   0.10 } $\\
PG~0804+761&
$  45.53 ^{+   0.22 }_{-   0.20 } $&
$  45.47 ^{+   0.04 }_{-   0.04 } $&
$  45.01 ^{+   0.02 }_{-   0.03 } $&
$  44.83 ^{+   0.04 }_{-   0.05 } $&
$  44.82 ^{+   0.27 }_{-   0.22 } $&
$  44.53 ^{+   0.10 }_{-   0.10 } $&
$  46.27 ^{+   0.14 }_{-   0.12 } $\\
H1039-074&
$  46.02 ^{+   0.01 }_{-   0.01 } $&
$  45.96 ^{+   0.01 }_{-   0.01 } $&
$  45.72 ^{+   0.02 }_{-   0.02 } $&
$  45.57 ^{+   0.02 }_{-   0.03 } $&
  ...&
  ...&
$  47.38 ^{+   0.11 }_{-   0.04 } $\\
NGC 3783&
$  43.88 ^{+   0.01 }_{-   0.01 } $&
$  44.02 ^{+   0.01 }_{-   0.01 } $&
$  43.82 ^{+   0.02 }_{-   0.02 } $&
$  43.20 ^{+   0.02 }_{-   0.02 } $&
$  42.95 ^{+   0.11 }_{-   0.08 } $&
$  43.18 ^{+   0.10 }_{-   0.10 } $&
$  44.86 ^{+   0.02 }_{-   0.02 } $\\
TON 1542&
$  44.99 ^{+   0.02 }_{-   0.03 } $&
$  44.84 ^{+   0.03 }_{-   0.03 } $&
$  44.47 ^{+   0.04 }_{-   0.06 } $&
$  44.39 ^{+   0.09 }_{-   0.10 } $&
$  44.62 ^{+   0.05 }_{-   0.05 } $&
$  43.90 ^{+   0.05 }_{-   0.05 } $&
$  45.75 ^{+   0.07 }_{-   0.07 } $\\
IRAS 13218+0552&
$  44.25 ^{+   0.06 }_{-   0.01 } $&
$  44.11 ^{+   0.07 }_{-   0.02 } $&
$  44.33 ^{+   0.17 }_{-   0.11 } $&
$  44.68 ^{+   0.08 }_{-   0.03 } $&
$  46.01 ^{+   0.00 }_{-   0.00 } $&
$  45.75 ^{+   0.00 }_{-   0.00 } $&
$  46.74 ^{+   0.13 }_{-   0.15 } $\\
MCG  -6-30-15 &
$ <  42.87 $&
$  42.73 ^{+   0.02 }_{-   0.68 } $&
$  43.24 ^{+   0.01 }_{-   0.01 } $&
$  43.23 ^{+   0.02 }_{-   0.02 } $&
$  43.39 ^{+   0.21 }_{-   0.21 } $&
$  42.87 ^{+   0.04 }_{-   0.04 } $&
$  44.29 ^{+   0.10 }_{-   0.27 } $\\
IC 4329A&
$  43.27 ^{+   0.01 }_{-   0.01 } $&
$  43.29 ^{+   0.01 }_{-   0.01 } $&
$  43.83 ^{+   0.01 }_{-   0.01 } $&
$  43.96 ^{+   0.02 }_{-   0.02 } $&
$  43.73 ^{+   0.04 }_{-   0.04 } $&
$  43.75 ^{+   0.09 }_{-   0.10 } $&
$  45.05 ^{+   0.04 }_{-   0.12 } $\\
H1419+480&
$  44.06 ^{+   0.10 }_{-   0.06 } $&
$  44.27 ^{+   0.10 }_{-   0.05 } $&
$  44.33 ^{+   0.13 }_{-   0.07 } $&
$  44.32 ^{+   0.15 }_{-   0.08 } $&
$  44.05 ^{+   0.05 }_{-   0.05 } $&
$  44.79 ^{+   0.03 }_{-   0.04 } $&
$  45.59 ^{+   0.10 }_{-   0.07 } $\\
Mkn 478 &
$  45.11 ^{+   0.10 }_{-   0.12 } $&
$  44.80 ^{+   0.03 }_{-   0.03 } $&
$  44.64 ^{+   0.02 }_{-   0.03 } $&
$  44.63 ^{+   0.04 }_{-   0.05 } $&
$  44.75 ^{+   0.16 }_{-   0.16 } $&
$  44.27 ^{+   0.06 }_{-   0.08 } $&
$  45.98 ^{+   0.09 }_{-   0.10 } $\\
H1537+339&
$  45.01 ^{+   0.26 }_{-   0.23 } $&
$  45.19 ^{+   0.13 }_{-   0.12 } $&
$  45.22 ^{+   0.07 }_{-   0.08 } $&
$  45.12 ^{+   0.09 }_{-   0.12 } $&
$  44.93 ^{+   0.50 }_{-   0.32 } $&
$  46.49 ^{+   0.00 }_{-   0.00 } $&
$  46.72 ^{+   0.08 }_{-   0.15 } $\\
Kaz 102&
$  45.53 ^{+   0.03 }_{-   0.03 } $&
$  45.33 ^{+   0.02 }_{-   0.02 } $&
$  44.87 ^{+   0.06 }_{-   0.06 } $&
$  44.66 ^{+   0.08 }_{-   0.07 } $&
$  44.37 ^{+   0.85 }_{-   0.52 } $&
 ...&
$  46.14 ^{+   0.16 }_{-   0.12 } $\\
E1821+643&
$  46.22 ^{+   0.12 }_{-   0.12 } $&
$  46.43 ^{+   0.03 }_{-   0.03 } $&
$  46.18 ^{+   0.01 }_{-   0.01 } $&
$  46.15 ^{+   0.01 }_{-   0.01 } $&
$  45.85 ^{+   0.31 }_{-   0.26 } $&
$  45.86 ^{+   0.02 }_{-   0.02 } $&
$  47.46 ^{+   0.00 }_{-   0.35 } $\\
H1834-653&
$  42.11 ^{+   0.02 }_{-   0.02 } $&
$  42.07 ^{+   0.01 }_{-   0.01 } $&
$  43.04 ^{+   0.01 }_{-   0.02 } $&
$  42.96 ^{+   0.01 }_{-   0.01 } $&
$  43.40 ^{+   0.34 }_{-   0.30 } $&
$  43.10 ^{+   0.08 }_{-   0.09 } $&
$  44.60 ^{+   0.04 }_{-   0.11 } $\\
MKN 509&
$  45.20 ^{+   0.04 }_{-   0.04 } $&
$  45.08 ^{+   0.03 }_{-   0.03 } $&
$  44.56 ^{+   0.01 }_{-   0.01 } $&
$  44.31 ^{+   0.05 }_{-   0.05 } $&
$  44.89 ^{+   0.06 }_{-   0.06 } $&
$  44.38 ^{+   0.10 }_{-   0.10 } $&
$  45.92 ^{+   0.05 }_{-   0.06 } $\\
NGC 7213&
$  42.58 ^{+   0.01 }_{-   0.00 } $&
$  42.62 ^{+   0.01 }_{-   0.00 } $&
$  42.89 ^{+   0.01 }_{-   0.01 } $&
$  42.95 ^{+   0.01 }_{-   0.01 } $&
$  42.91 ^{+   0.06 }_{-   0.03 } $&
$  42.64 ^{+   0.15 }_{-   0.14 } $&
$  44.15 ^{+   0.04 }_{-   0.21 } $\\
MR2251-178&
$  45.16 ^{+   0.11 }_{-   0.11 } $&
$  45.00 ^{+   0.10 }_{-   0.10 } $&
$  44.75 ^{+   0.03 }_{-   0.03 } $&
$  44.66 ^{+   0.11 }_{-   0.11 } $&
$  44.52 ^{+   0.30 }_{-   0.27 } $&
$  44.54 ^{+   0.03 }_{-   0.03 } $&
$  45.88 ^{+   0.12 }_{-   0.11 } $\\
MCG -2-58-22 &
$  44.92 ^{+   0.07 }_{-   0.08 } $&
$  44.71 ^{+   0.07 }_{-   0.09 } $&
$  44.52 ^{+   0.14 }_{-   0.30 } $&
$  44.58 ^{+   0.11 }_{-   0.17 } $&
$  44.52 ^{+   0.27 }_{-   0.26 } $&
$  44.65 ^{+   0.09 }_{-   0.09 } $&
$  45.72 ^{+   0.12 }_{-   0.15 } $\\
\enddata
\tablenotetext{a}{Values are logarithm of luminosity in units of
erg~s$^{-1}$. Cosmological model with $q_{o} = 0$ and $H_{o}$ =
50~km~s$^{-1}$Mpc$^{-1}$ was used.}
\end{deluxetable}

\clearpage
\begin{deluxetable}{lrrc}
\tablenum{7}
\tablecaption{Dust temperature range}
\tablewidth{195pt}
\tablehead{
\colhead{Name} &
\colhead{$T_{min}$}&
\colhead{$T_{max}$}&
\colhead{$\alpha_{cut}$$^a$}
}
\startdata
MKN 1152&    $<$30 & 1000& \ldots\\
MKN 590\tablenotemark{b}&        20 & 1000& \ldots\\
ESO 198$-$G24 &   40 & 1000&0.2\\ 
3A 0557$-$383 &   20 & 1000&0.7\\
PG 0804+761&    40 & 1000&0.6\\
H 1039$-$074&      40 &  500&\ldots\\
NGC 3783&       30 & 1000&1.4\\
TON 1542&    $<$80 & 1000&\ldots\\
IRAS 13218+0552&50 & 1000&1.8\\
MCG $-$6$-$30$-$15 & 40 & 1000&1.3\\
IC 4329A&       40 & 1000&0.8\\
H 1419+480&     40 & 1000&0.3\\
MKN 478 &       40 & 1000&1.8\\
H 1537+339&   $<$40 &  500&\ldots\\
KAZ 102&        40 & 1000&0.2\\
E 1821+643&      40 & 1000&2.5\\
H 1834-653&     50 & 1000&1.6\\
MKN 509&        40 & 1000&2.2\\
NGC 7213\tablenotemark{b}&    $<$20 & 1000&\ldots\\
MR 2251$-$178&    40 & 1000&3.7\\
MCG $-$2$-$58$-$22 &  40 & 1000&2.5
\tablenotetext{a}{$\alpha_{cut}$ is the far-IR cutoff slope}
\tablenotetext{b}{the SED is unusually flat and covers full range
of observable temperatures; no strong constraints on temperature can be
placed.} 
\enddata
\end{deluxetable}

\begin{deluxetable}{lccccc}
\tablenum{8}
\tablecaption{Parameters for model SEDs which reproduce the observed AGN SEDs}
\tablewidth{0pt}
\tablehead{
\colhead{QSO Name} & 
\colhead{L$_{agn}$ (\lsun)}	& 
\colhead{$R_{min}$} &
\colhead{M$_{torus}$ (\msun)} & 	
\colhead{h$_0$(pc)} & 
\colhead{Inclination} 
}
\startdata
PG~0804+761	& $6.0*10^{12}$	&	5pc&3$*10^8$	&0.04	& 0$^{o}$\\
MR~2251$-$178	& $2.5*10^{12}$	&	2pc&8$*10^9$	&0.025	& 0$^{o}$\\
IR~13218+0552	& $5.0*10^{13}$	&	2pc&5$*10^9$	&0.1 	&65$^{o}$\\
\enddata
\end{deluxetable}

\begin{table}
\dummytable\label{tab:names}      
\end{table}

\begin{table}
\dummytable\label{tab:ISO}        
\end{table}

\begin{table}
\dummytable\label{tab:ebv_star}   
\end{table}

\begin{table}
\dummytable\label{tab:photom_jhk} 
\end{table}

\begin{table}
\dummytable\label{tab:integrIR}   
\end{table}

\begin{table}
\dummytable\label{tab:optuv}      
\end{table}

\begin{table}
\dummytable\label{temp:tb}        
\end{table}

\begin{table}
\dummytable\label{tb:pars}        
\end{table}

\end{document}